\documentclass[journal]{IEEEtran}
\usepackage[boxruled]{algorithm2e}
\usepackage{flushend}
\usepackage{cite}
\usepackage{graphicx}
\usepackage{psfrag}
\usepackage{subfigure}
\usepackage{url}
\usepackage{amsmath}
\usepackage{array}
\usepackage{amssymb}
\usepackage{amsfonts}
\newtheorem{proposition}{Proposition}
\newtheorem{lemma}{Lemma}

\newtheorem{corollary}{Corollary}

\newtheorem{example}{Example}

\begin{document}
\title{On network coding for acyclic networks with delays}
\author{
\authorblockN{K.~Prasad and B.~Sundar Rajan\\}
\authorblockA{Dept. of ECE, IISc, Bangalore 560012, India\\
Email: \{prasadk5,bsrajan\}@ece.iisc.ernet.in\\}
}
\date{\today}
\maketitle
\thispagestyle{empty}
\let\thefootnote\relax\footnotetext
{
Part of the content of this work has appeared in the Proceedings of IEEE Information Theory Workshop held at Paraty, Brazil, October 16-20, 2011.
}
\begin{abstract}
Problems related to network coding for acyclic, instantaneous networks (where the edges of the acyclic graph representing the network are assumed to have zero-delay) have been extensively dealt with in the recent past. The most prominent of these problems include $(a)$ the existence of network codes that achieve maximum rate of transmission, $(b)$ efficient network code constructions, and $(c)$ field size issues. In practice, however, networks have transmission delays. In network coding theory, such networks with transmission delays are generally abstracted by assuming that their edges have integer delays. Note that using enough memory at the nodes of an acyclic network with integer delays can effectively simulate instantaneous behavior, which is probably why only acyclic instantaneous networks have been primarily focused on thus far. In this work, we elaborate on issues ($(a), (b)$ and $(c)$ above) related to network coding for acyclic networks with integer delays, which have till now mostly been overlooked. We show that the delays associated with the edges of the network cannot be ignored, and in fact turn out to be advantageous, disadvantageous or immaterial, depending on the topology of the network and the problem considered i.e., $(a), (b)$ or $(c).$ In the process, we also show that for a single source multicast problem in acyclic networks (instantaneous and with delays), the network coding operations at each node can simply be limited to storing old symbols and coding them over a binary field. Therefore, operations over elements of larger fields are unnecessary in the network, the trade-off being that enough memory exists at the nodes and at the sinks, and that the sinks have more processing power.
\end{abstract}
\section{Introduction}
\label{sec1}

Network coding was introduced in \cite{ACLY} as a means to improve the rate of transmission in networks. Linear network coding was introduced in \cite{CLY} and it was found to be sufficient to achieve the maxflow-mincut capacity in certain scenarios such as multicast. The linear network coding problem on a network with given sink demands can be considered to have three major subproblems. 
\begin{itemize}
\item Existence of a network code that satisfies the demands.
\item Efficient construction of such a network code. 
\item Minimum field size for the existence of such a network code.
\end{itemize}

An algebraic theory of network coding was developed in \cite{KoM}, which converted the existence problem of network coding into an algebraic geometry problem.  As for the latter two, most of the literature in network coding has focused on the multicast problem, i.e., where all sinks demand the information generated by all the sources in the network. A polynomial-time algorithm for designing a multicast network code on a single-source acyclic instantaneous (zero-delay) network was presented in \cite{JSCEEJT}. This algorithm was further generalized in \cite{ErF} and \cite{BaY1} for the case of multicast on networks with cycles. Note that the notion of delays in the network is inherent to any algorithm on cyclic networks. Delays are therefore assumed either on the edges of the networks that contribute to the cycles alone, or throughout the network. An information flow decomposition based approach to the problem of  network code construction was discussed in \cite{FrS}.

Computing the minimum field size required to solve a network coding problem is known \cite{LeL1} to be NP-hard. However, for the multicast case on acyclic networks and certain kinds of cyclic networks, it is known \cite{KoM} \cite{JSCEEJT} that a field size larger than the number of sinks in the network is sufficient. Further results on the field size issue can be found in \cite{BaY1}\cite{FrS} \cite{TFR,CFS,LeL2}. In certain networks, linear network coding itself is found to be insufficient to achieve the given demands \cite{DFZ}.

The case of acyclic networks with delays was abstracted in \cite{KoM} as acyclic networks where each edge in the network has an integer delay associated with it. With this setting, the authors of \cite{KoM} were able to naturally generalize the notion of linear network coding and thereby the framework for the problem of the existence of a linear network code on such networks was presented. According to the framework of \cite{KoM}, a network code on an acyclic network with general demands has to satisfy two conditions at every sink to be a solution for the network, which we refer to as $(a)$ \textit{invertibility} conditions, which have to be satisfied to recover the information sequences demanded at each sink, and (b) \textit{zero-interference} conditions, which have to be satisfied so that information sequences not needed at a sink do not interference with those that are demanded (a formal description of these conditions are given in Section \ref{sec3}). If at least one such network coding solution exists for the network, then the network is said to be \textit{solvable}.
 
A \textit{delay profile} for a network consists of a set of non-negative integers, one for each edge in the network indicating the integer delay experienced by the symbols on that edge. For a given delay profile for a network, it has been noted (see \cite{WZY}\cite{PrR1}, for example) that an instantaneous behaviour can be simulated in acyclic networks with integer delays using enough memory at the nodes of the network. It is assumed in most of network coding literature that this can always be done, and that this is indeed the source of the instantaneous behaviour in the network. However, several questions remain unexplored, such as how solvability, field size, etc., are affected when there is a change in the delay profile of the network, given that the network has already been configured to be instantaneous under a known delay profile. 

In \cite{SJL}, it is shown that for multicast networks which are equipped with memory at the nodes, there always exists a network code (using memory at the nodes) which is a valid solution for the network for any delay profile under any field size. The authors of \cite{SJL} further show that such a \textit{delay-invariant} code can be found with high probability using a random choice of the network coding coefficients from a large field. They also give a deterministic algorithm to construct such a delay-invariant multicast network code for acyclic and cyclic networks, as long as the field size is larger than the number of sinks. In \cite{Eff}, it is shown that under the conditions where the nodes in the network are always equipped with enough memory to counteract any amount of delays in the network, a network (with arbitrary demands) is solvable for any delay profile if and only if it is solvable with an all-zero delay profile. In other words, networks codes which are solutions for the network under a certain delay profile can always be converted into solutions for the network under a different delay profile by utilizing the appropriate number of memory elements at the nodes of the network. 

In this work, we wish to study the effect of the relationship between some of the network coding problems for instantaneous networks and their counterparts with non-trivial delay profiles. The instantaneous network of a network $\cal G$, which is referred to as ${\cal G}_{inst}$ throughout the paper, corresponds to the network $\cal G$ with a known delay profile where memory elements have already been used to simulate instantaneous (zero-delay) behaviour. We assume that any delay profile in $\cal G$ can only have greater delays (on one or many edges) compared to those in the basic delay profile that gives rise to the instantaneous network ${\cal G}_{inst}.$ Therefore, throughout the paper, we view the instantaneous networks as networks with the all-zero delay profile.

Many of the results in this paper compare the network coding problems on ${\cal G}_{inst}$ with an \textit{unit-delay network}, ${\cal G}_{ud}.$ The unit-delay network is the network where the delays in the edges are exactly one unit above the delays in ${\cal G}_{inst}.$  To derive these comparison results, we assume that the intermediate (non-source non-sink) nodes of $\cal G$ are equipped with memory sufficient only to simulate the instantaneous behaviour of ${\cal G}_{inst}$. This forms the major difference between our results and that of \cite{Eff}, where there is no bound on the amount of memory used by the intermediate nodes. In contrast with \cite{SJL,Eff}, we concern ourselves with analysing whether delays over those in ${\cal G}_{inst}$ can be advantageous or disadvantageous or neutral, given that the intermediate nodes do not have memory beyond what is used by them to simulate instantaneous behaviour in ${\cal G}_{inst}.$ Following our framework of viewing ${\cal G}_{inst}$ as ${\cal G}$ with the all-zero delay profile, and because of the fact that we limit the amount of memory elements at the intermediate nodes of ${\cal G}$, we refer to the intermediate nodes of $\cal G$ as being \textit{memory-free}, i.e, utilizing no memory elements. The unit-delay network ${\cal G}_{ud}$ is then appropriately viewed as the network $\cal G$ with an all-one delay profile (i.e, one where all edges have a delay of unity associated with them). Although our results show the comparisons between network coding on ${\cal G}_{inst}$ and ${\cal G}_{ud},$ they can be generalized without much difficulty to general delay profiles. The all-one delay profile is chosen only because it is sufficient to illustrate the differences obtained through our results between instantaneous networks and those with delays, and less cumbersome to handle in terms of notation.

Table \ref{tab1} summarises the relationships obtained in this paper between some of the network coding problems for instantaneous and unit-delay networks. 
\begin{table*}
\normalsize
\centering
\caption{Relationship between network coding problems between ${\cal G}_{inst}$ and ${\cal G}_{ud}$ for an acyclic graph ${\cal G}.$ }
\begin{tabular}{|c|c|c|}
\hline
\textbf{Property of interest} & \textbf{If the property holds for}  $\boldsymbol{{\cal G}_{inst},}$ & \textbf{If the property holds for}  $\boldsymbol{{\cal G}_{ud},}$ \\
&\textbf{does it continue to hold for }$\boldsymbol{{\cal G}_{ud}}$\textbf{?} & \textbf{does it continue to hold for }$\boldsymbol{{\cal G}_{inst}}$\textbf{?} \\
\hline
Solvability & \textit{Yes}, if zero-interference conditions & \textit{Yes} (Proposition \ref{prop2}). \\
&are satisfied (Proposition \ref{prop1}),  & \\
& or for multicast (Corollary \ref{cor1}). &\\ 
& \textit{No}, if they are not satisfied (Example \ref{exm1}). &\\
\hline 
Non-solvability & \textit{Yes} (Corollary \ref{cornonsolvability}). & \textit{No} (Example \ref{exm1}).  \\
\hline
Polynomial-time algorithms & \textit{No}, in general, as ${\cal G}_{ud}$ might not & \textit{Yes} (Corollary \ref{corollaryconstruction}). \\
for code construction  & even be solvable (Example \ref{exm1}). &   \\
\hline
Solvability over &\textit{No}, in general, as ${\cal G}_{ud}$ might not & \textit{No}, in general (Corollary \ref{cor4}), as illustrated \\
a particular field &  even be solvable (Example \ref{exm1}). & in Example \ref{exm2} and Example \ref{exm3}.\\
&\textit{Yes}, for certain conditions on & \textit{Yes}, under certain conditions  \\
&  topology given by Proposition \ref{prop3}, & given by Proposition \ref{prop3}.\\
& or for multicast (Corollary \ref{cor1}). & \\
\hline
Non-solvability over & \textit{No}, in general (Corollary \ref{cor4}), as illustrated & \textit{No}, in general (Example \ref{exm1}). \\
a particular field & in Example \ref{exm2} and Example \ref{exm3}. & \textit{Yes}, under conditions on topology  \\
 & & given by Corollary \ref{cor3}.\\
\hline
\end{tabular}
\label{tab1}
\end{table*}
The contributions of our work are as follows.
\begin{itemize}
\item We prove that the solvability of ${\cal G}_{inst}$ preserves the invertibility conditions (Proposition \ref{prop1}) in ${\cal G}_{ud}$, but not necessarily zero-interference conditions (Example \ref{exm1}). On the other hand, we prove that if ${\cal G}_{ud}$ is solvable, then ${\cal G}_{inst}$ is always solvable (Proposition \ref{prop2}), and thereby proving that if ${\cal G}_{inst}$ is not solvable, then so is ${\cal G}_{ud}$ (Corollary \ref{cornonsolvability}). These results on the relationship between the solvability and non-solvability of ${\cal G}_{inst}$ and ${\cal G}_{ud}$ are tabulated in the first two rows of Table \ref{tab1}. 
\item We show that whenever there is a polynomial-time algorithm for constructing a network code for ${\cal G}_{ud}$ that satisfies all sink demands, then there is a polynomial-time algorithm for constructing a network code for ${\cal G}_{inst}$ which satisfies all sink demands (Corollary \ref{corollaryconstruction}). The third row of Table \ref{tab1} captures these results.
\item We prove that under certain conditions on the topology of the network there exists an equivalence between a network code over any particular field constructed on ${\cal G}_{inst}$ and ${\cal G}_{ud}$ (Proposition \ref{prop3}). Thus, for networks obeying the constraints given in Proposition \ref{prop3}, the minimum field size for constructing a network code satisfying all demands for ${\cal G}_{inst}$ and ${\cal G}_{ud}$ is the same. We also prove that under such constraints on topology, the non-solvability of ${\cal G}_{ud}$ implies the non-solvability of ${\cal G}_{inst}$ (Corollary \ref{cor3}). The last two rows of Table \ref{tab1} lists these results.
\item We prove that there exist networks for which the delays prove useful for the field size problem, i.e., network codes can be constructed over a smaller field size for ${\cal G}_{ud}$ compared to ${\cal G}_{inst},$ and also show a construction of such networks (Corollary \ref{cor4}). These results are also tabulated in the last two rows of Table \ref{tab1}. Towards that end, we prove the feasibility of two multicast algorithms, one of which works for acyclic networks (instantaneous and with delays) and was conjectured in \cite{BaY2} based on the multicast algorithm of \cite{JSCEEJT}, and the other works for certain special acyclic networks (Proposition \ref{prop4} and Proposition \ref{prop5}). These modified algorithms employ low-complexity encoding at the intermediate nodes over $\mathbb{F}_2$ using memory elements, while possibly demanding a larger complexity of decoding at the sinks compared to traditional network coding schemes.
\end{itemize}

The rest of this work is organized as follows. In Section \ref{sec2}, we set up the model and the terminology for acyclic networks with delays. In Section \ref{sec3}, we explore the relationship between the network code existence problem in ${\cal G}_{ud}$ and ${\cal G}_{inst}$ for an acyclic network ${\cal G}$ with given set of demands, and also present examples where having delays prevent the existence of any solution for ${\cal G}_{ud}$ while solutions exist for ${\cal G}_{inst}.$ In Subsection \ref{subsec4b}, we analyze the conditions on topology which result in an equivalence of network coding solutions between  ${\cal G}_{ud}$ and ${\cal G}_{inst}.$  After briefly reviewing the Linear Information Flow (LIF) algorithm of \cite{JSCEEJT} in Subsection \ref{lifalg}, in Subsection \ref{delayandcode} we use a modified version of this algorithm to obtain a class of networks in which delays prove beneficial in the minimum field size problem, i.e., where feasible binary network codes always exist for ${\cal G}_{ud}$ irrespective of the field size required for ${\cal G}_{inst}.$ We conclude the paper in Section \ref{sec5} with remarks and directions for further research.
\section{Network codes for acyclic networks with delays}
\label{sec2}
Following the terminology of \cite{JSCEEJT}, an acyclic network is modeled as an acyclic graph $\cal G$ with $\cal V$ being the set of nodes and $\cal E$ the set of edges in the network. The set ${\cal V}$ contains a set of source nodes $\cal S$ and a set of sink nodes $\cal T.$ We assume that the sources have no incoming edges in the network, while the sinks have no outgoing edges. The time unit under consideration shall imply one use of the channels in the network. Each source $s \in {\cal S}$ generates $h_s$ information sequences at the rate of $h_s$ $\mathbb{F}_q$ symbols per every time unit, $\mathbb{F}_q$ being the finite field with $q$ symbols. For each source $s \in {\cal S},$ we introduce $h_s$ parallel edges (denoted by ${\cal E}_s$) incoming at $s,$ which carry the $h_s$ information sequences to the source $s.$ Let $h=\sum_{s\in \cal S}h_s.$ 

Assuming an ordering on the set of information sequences available at the sources, let ${\cal I}_t$ denote an indicator function for a sink $t\in \cal T,$ defined as  
\begin{equation*}
{\cal I}_t:\left\{1,2,...,h\right\} \rightarrow \left\{0,1\right\},
\end{equation*}
such that, ${\cal I}_t(i)=1,$ if sink $t$ demands the $i^{th}$ information sequence, and $0$ otherwise. Let $\cal C$ denote the collection of the functions ${\cal I}_t, \forall  t \in {\cal T}.$

Each sink node $t\in \cal T$ demands some subset of size $h_t$ of the $h$ information sequences generated at the sources. Let $h_{_{\cal T}}=\sum_{t\in{\cal T}}h_t.$ For each sink $t,$ we assume $h_t$ imaginary outgoing edges from $t,$ denoted by ${\cal E}_t.$ We represent a network ${\cal G}({\cal V},{\cal E})$ with a set of sources ${\cal S}$ and a set of sinks ${\cal T}$ with a set of demands given by ${\cal C}$ as ${\cal G}({\cal V},{\cal E},{\cal S},{\cal T},{\cal C}).$

Every edge in the directed graph representing the network has a capacity of one $\mathbb{F}_q$ symbol. We abstract the case of networks with delay by assuming a unit-delay associated with edges of the graph $\cal G$, represented by the parameter~$z$. We denote the graph ${\cal G}({\cal V},{\cal E})$ along with the delays as ${\cal G}_{ud},$ the unit-delay version of $\cal G$ or simply the \textit{unit-delay network} ${\cal G}_{ud}$. Note that network links with integer delays greater than unit are modeled as serially concatenated edges in the directed multi-graph. Because of this reason, we view networks with integer delays and those with unit-delays equivalently.

The set of symbols generated at the sources at any particular instant of time is called a \textit{generation} of symbols. Any node in a unit-delay network may receive information of different generations on its incoming edges at any particular time instant. Except for the discussion in Subsection \ref{delayandcode}, we assume that the intermediate (non-sink, non-source) nodes are memory-free and merely transmit a $\mathbb{F}_q$ linear combination of the incoming sequences on their outgoing edges. Also, the zero-delay version of ${\cal G},$ referred to as the \textit{instantaneous network}, is denoted by ${\cal G}_{inst}.$ 
The following notations will be used throughout the paper. 

\small
\begin{align}
\nonumber
&\Gamma_I(v)&:&~\text{Set of incoming (including imaginary) edges at node}~v \\
\nonumber
&\Gamma_O(v)&:&~\text{Set of outgoing (including imaginary) edges at node}~v \\
\nonumber
&\delta_I(v)&:&~|\Gamma_I(v)|.\\
\nonumber
&\delta_O(v)&:&~|\Gamma_O(v)|.\\
\nonumber
&v\hspace{-0.1cm}=\hspace{-0.1cm}head(e)\hspace{-0.3cm}&:&~\text{if}~e\in\Gamma_I(v).\\
\nonumber
&v\hspace{-0.1cm}=\hspace{-0.1cm}tail(e)\hspace{-0.3cm}&:&~\text{if}~e\in\Gamma_O(v).
\end{align}
\normalsize

For an edge $e \in {\cal E}\cup {\cal E}_t,$ we define the \textit{local encoding vector} as a $\delta_I(tail(e))$-length vector, $\left(m_{e,p}(z):p\in\Gamma_I(tail(e))\right),$
where $m_{e,p}(z)\in\mathbb{F}_q(z),$ the field of rational functions over $\mathbb{F}_q.$ The local encoding vector determines the sequence $y_e(z)=\sum_i y_{e,i}z^i \left(y_{e,i} \in \mathbb{F}_q\right.$ being the symbol at $i^{th}$ time index$\left.\right)$ flowing on edge $e$ based on the sequences incoming at $tail(e),$ i.e., 
\begin{equation}
\label{eqn4}
y_e(z)=\sum_{p\in\Gamma_I(tail(e))}m_{e,p}(z)y_p(z).
\end{equation}
Note that as the intermediate nodes are allowed to take only $\mathbb{F}_q$ linear combinations of the incoming sequences, we have for an edge $e \notin \Gamma_O(s)$ (for any $s\in \cal S$), $m_{e,p}(z)=zm_{e,p},$ for some $m_{e,p}\in \mathbb{F}_q$ and the parameter $z$ denotes the delay incurred during the transmission through edge $e.$ For an edge $e \in \Gamma_O(s)$ of some source $s\in \cal S,$ we have $m_{e,p}(z)=z\tilde{m}_{e,p}(z),$ for some $\tilde{m}_{e,p}(z)\in \mathbb{F}_q(z),$ as we let the sources take arbitrary combinations over $\mathbb{F}_q(z).$ The additional $z$ again denotes the delay incurred on the edge $e.$ For ${\cal G}_{inst},$ note that 
\begin{equation}
\label{eqn5}
m_{e,p}(z)=m_{e,p}\in \mathbb{F}_q,
\end{equation}
for any pair of edges $e$ and $p,$ and therefore the corresponding input-output relationship for any edge $e$ is given independent of the time index as
\begin{equation*}
y_e=\sum_{p\in\Gamma_I(tail(e))}m_{e,p}y_p,
\end{equation*}
where $y_e,y_p \in \mathbb{F}_q.$ 

Let $\boldsymbol{m}$ denote the set of all \textit{local encoding coefficients} (all taking values from $\mathbb{F}_q$). For ${\cal G}_{ud},$ $\boldsymbol{m}$ is the set of all $\mathbb{F}_q$ coefficients of the numerators and denominators of all $m_{e,p}(z)$. For ${\cal G}_{inst},$  $\boldsymbol{m}$ denotes the set of all $m_{e,p}.$ The difference between the two will be clear from the context.

The network coding problem implies a choice of the local encoding coefficients $m_{e,p}$ such that each sink can recover the information it demands. Because of the linearity of (\ref{eqn4}), we can associate with every edge $e$ a $h$-length \textit{global encoding vector} over $\mathbb{F}_q(z).$ The global encoding vector $\boldsymbol{b}(e)$ of edge $e,$ indicates the particular $\mathbb{F}_q(z)$ linear combination of the $h$ information sequences, flowing in $e.$ The global encoding vectors of the $h$ incoming edges at the sources correspond to the basis vectors of ${\mathbb F}_q^h$. By (\ref{eqn4}), the vector $\boldsymbol{b}(e)$ can be recursively calculated from the global encoding vectors of the edges incoming at $tail(e).$ The global encoding vectors are well defined because of the acyclicity of the network.

Having ordered the $h$ input sequences and the $h_{_{\cal T}}$ output sequences, the input-output relationship of  ${\cal G}_{ud}$ can be represented as a $h \times h_{_{\cal T}}$ matrix over $\mathbb{F}_q(z)$ called the \textit{overall transfer matrix} \cite{KoM}, $M(z),$ of the network, the columns of which are the global encoding vectors of the imaginary outgoing edges from the sinks. The transfer matrix corresponding to a particular sink $t,$ is the $h \times h_t$ matrix $M_t(z),$ the columns of which are the global encoding vectors of the imaginary outgoing edges from the sink $t.$ Therefore, for $\boldsymbol{x}(z)$ being the $h$-length input vector and $\boldsymbol{y_t}(z)$ being the $h_t$-length output vector at sink $t,$ we have $\boldsymbol{y_t}(z)=\boldsymbol{x}(z)M_t(z).$ For ${\cal G}_{inst},$ the components of the global encoding vectors and network transfer matrices are all elements from $\mathbb{F}_q.$ For more details on the structure of these matrices, we refer the reader to \cite{KoM}. 
\section{Existence of Network codes for acyclic network with delays}
\label{sec3}
The problem of network code existence was presented from an algebraic geometry point of view in \cite{KoM}. The local encoding coefficients $\boldsymbol{m}$ are assumed to be variables which can take values from a large enough finite field. A network code, i.e., a particular choice of the set of all local encoding coefficients $\boldsymbol{m},$ is defined to be \textit{feasible}, i.e., it achieves the given set of demands at the sinks, if the following two conditions are satisfied.
\begin{itemize}
\item \textit{Invertibility conditions:} For each sink $t,$ the $h_t \times h_t$ submatrix $M'_t(z)$ of $M_t(z),$ the rows of which corresponding to the inputs demanded at sink $t,$ is invertible over $\mathbb{F}_q(z).$ 
\item \textit{Zero-Interference conditions:} For each sink $t,$ the elements of the matrix $M_t(z)$ which are not part of $M'_t(z)$ are zero. 
\end{itemize}

Note that if the mincut between any source $s$ and any sink $t$ is less than the number of information sequences demanded by $t$ from $s,$ then the network coding problem is clearly not solvable. Besides the mincut conditions, the topology of the network also affects the ability to satisfy the demands in the network. 

For each sink $t,$ some elements of $M_t(z)$ are not a part of the $M'_t(z)$ matrix. Let $f_1,f_2,...,f_K$ be all such elements, for all possible sinks $t\in \cal T.$ Note that each $f_i \in \mathbb{F}_q(z)$ for any particular choice of $\boldsymbol{m}$, hence we represent each $f_i$ as $f_i(\boldsymbol{m},z).$ Similarly, let $g_1(\boldsymbol{m},z),g_2(\boldsymbol{m},z),...,g_L(\boldsymbol{m},z)$ be the determinants of the $M'_t(z)$ matrices. Let $g(\boldsymbol{m},z)=\prod_{i=1}^{L}g_i(\boldsymbol{m},z).$ The invertibility and zero-interference conditions then imply that the assignment of $\boldsymbol{m}$ should satisfy $g(\boldsymbol{m},z)\neq 0$ and $f_1(\boldsymbol{m},z)=f_2(\boldsymbol{m},z)=...=f_K(\boldsymbol{m},z)=0$ respectively. Similar conditions (except for the delay parameter $z$) for feasibility hold good for the ${\cal G}_{inst}$ also. Note that for certain network topologies or sink demands, the invertibility conditions alone will suffice for feasibility, while the zero-interference conditions might not arise at all \cite{KoM}. The multicast case, where all sinks demand all the information sequences, is one such example.

We now provide some results regarding the question of whether the solvability of ${\cal G}_{ud}$ implies the solvability of ${\cal G}_{inst}$ also, and vice versa. The following proposition is a generalized version of Proposition 1 of \cite{PrR}, where the statement was proved only for a multicast case. A simpler proof for this proposition can also be derived easily from the results of \cite{SJL}.
\begin{proposition}
\label{prop1}
Let ${\cal G}_{ud}({\cal V},{\cal E},{\cal S},{\cal T},{\cal C})$ be an acyclic, unit-delay network with a given set of sink demands and ${\cal G}_{inst}({\cal V},{\cal E},{\cal S},{\cal T},{\cal C})$ be the corresponding instantaneous network. Let $\boldsymbol{m'}$ be a set of local encoding kernels which result in a network code for ${\cal G}_{inst}$, satisfying the invertibility conditions. Then $\boldsymbol{m'}$ continues to satisfy the invertibility conditions for ${\cal G}_{ud}.$
\end{proposition}	
\textit{Proof:} See Appendix \ref{prop1proof}.

For the multicast case, which has no zero-interference conditions, we then have the following corollary, which was proved in \cite{PrR}.
\begin{corollary}
\label{cor1}
Let ${\cal G}_{ud}({\cal V},{\cal E},{\cal S},{\cal T},{\cal C})$ be an acyclic, unit-delay network with multicast demands, i.e., all sinks require all the information sequences, and ${\cal G}_{inst}({\cal V},{\cal E},{\cal S},{\cal T},{\cal C})$ be the corresponding instantaneous network. Then a feasible network code for ${\cal G}_{inst}$ continues to be feasible for ${\cal G}_{ud}.$ 
\end{corollary}

In a general non-multicast network coding problem, it might not be possible to satisfy the zero-interference conditions in the network ${\cal G}_{ud},$ though they can be satisfied in the network ${\cal G}_{inst}.$  This is because of the fact that different flows which cancelled out the interference in ${\cal G}_{inst}$ can take paths of different delays in the corresponding acyclic network with delays, thereby preventing the cancelling effect. Example \ref{exm1} illustrates one such network, for which there exists solutions in ${\cal G}_{inst},$ but none for ${\cal G}_{ud}.$
\begin{example}
\label{exm1}
Consider the network ${\cal G}$ shown in Fig. \ref{fig1}. Let the field under consideration be $\mathbb{F}_q.$ Source $s_1$ has a sequence $x_1(z)$, which has to be conveyed to sink $t_1,$ while the sequence $x_2(z)$ at source $s_2$ has to be conveyed to sink $t_2.$ In both ${\cal G}_{inst}$ and ${\cal G}_{ud},$ the topology of the network demands that the linear combination of the two incoming sequences at node $v_1$ should be such that both the local encoding coefficients are non-zero. 

In ${\cal G}_{inst},$ the information sequence $x_1(z)$ is cancelled out at node $v_2$ to enable sink $t_2$ to receive $x_2(z),$ and similarly cancellation of $x_2(z)$ happens at node $v_3$ for sink $t_1.$ In ${\cal G}_{ud},$ this cancellation, while being necessary for the network code to be feasible, cannot happen at the nodes $v_2$ and $v_3$ because of the disparity in the delays of the flows at their incoming edges. Since the choice of our finite field was arbitrary, it is therefore clear that unless memory is used at some of the intermediate nodes, there exists no feasible network code for this network considered with delays over any finite field.
\begin{figure}[htbp]
\centering
\includegraphics[totalheight=2.5in,width=2.5in]{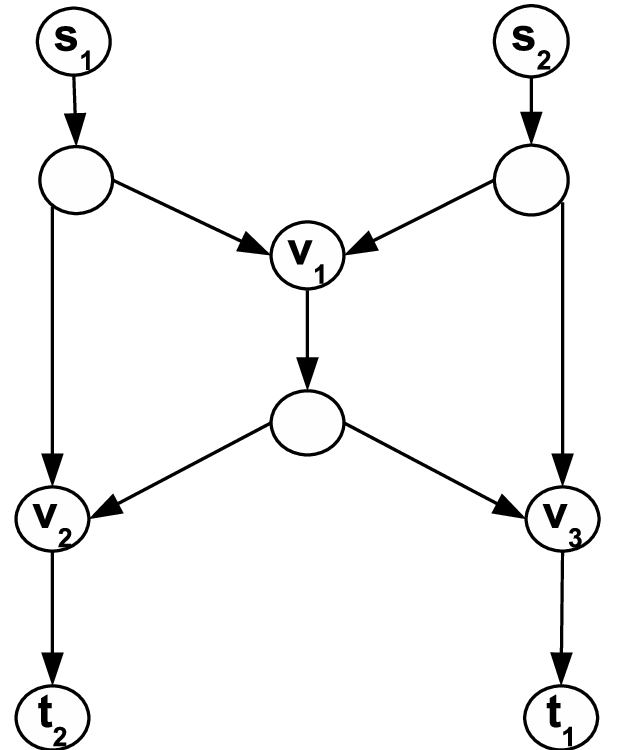}
\caption{A network $\cal G$ where zero-interference conditions fail to hold in ${\cal G}_{ud}$}	
\label{fig1}	
\end{figure}
\end{example}

In light of Proposition \ref{prop1} and Example \ref{exm1}, it can be observed that the solvability of a  network coding problem for ${\cal G}_{inst}$ need not imply solvability for ${\cal G}_{ud}.$ The following proposition answers the reverse problem, i.e., that solvability of a given ${\cal G}({\cal V},{\cal E},{\cal S},{\cal T},{\cal C})$ for ${\cal G}_{ud}$ always implies its solvability for ${\cal G}_{inst}.$
\begin{proposition}
\label{prop2}
Let ${\cal G}_{ud}({\cal V},{\cal E},{\cal S},{\cal T},{\cal C})$ be an acyclic, unit-delay network with a given set of sink demands and ${\cal G}_{inst}$ be the corresponding instantaneous network. If there exists a feasible network code for ${\cal G}_{ud}({\cal V},{\cal E},{\cal S},{\cal T},{\cal C}),$ then there exists a feasible network code for ${\cal G}_{inst}({\cal V},{\cal E},{\cal S},{\cal T},{\cal C}).$
\end{proposition}
\textit{Proof:} See Appendix \ref{prop2proof}.

Proposition \ref{prop2} leads to the following corollary.
\begin{corollary}
\label{cornonsolvability}
Let ${\cal G}_{ud}({\cal V},{\cal E},{\cal S},{\cal T},{\cal C})$ be an acyclic, unit-delay network with a given set of sink demands and ${\cal G}_{inst}$ be the corresponding instantaneous network. If there exists no feasible network code for ${\cal G}_{inst}({\cal V},{\cal E},{\cal S},{\cal T},{\cal C}),$ then there exists no feasible network code for ${\cal G}_{ud}({\cal V},{\cal E},{\cal S},{\cal T},{\cal C}).$
\end{corollary}

Note that the proof of Proposition \ref{prop2} involved an actual construction of a feasible network code for ${\cal G}_{inst}$ starting from a feasible network code for ${\cal G}_{ud}.$ Such a construction implies the following corollary on a polynomial-time construction for a feasible network coding solution for ${\cal G}_{inst}.$  
\begin{corollary}
\label{corollaryconstruction}
Let ${\cal G}_{ud}({\cal V},{\cal E},{\cal S},{\cal T},{\cal C})$ be an acyclic, unit-delay network with a given set of sink demands and ${\cal G}_{inst}$ be the corresponding instantaneous network. If there exists a polynomial-time construction algorithm for a feasible network coding solution on ${\cal G}_{ud},$ then there exists a polynomial-time construction algorithm for a feasible network coding solution on ${\cal G}_{inst}.$
\end{corollary}
\textit{Proof:} See Appendix \ref{corollaryconstructionproof}.

\section{Relationship between the minimum field size problem for ${\cal G}_{ud}$ and ${\cal G}_{inst}$}
\label{sec4}
In this section, we discuss the effect of considering delays in the network on the field size over which a valid network code can be designed for an acyclic network ${\cal G}.$ We assume that ${\cal G}_{ud}$ is solvable, which mean that ${\cal G}_{inst}$ is also solvable, according to Proposition \ref{prop2}. Note that Proposition \ref{prop1} already gives a small insight into the field size issue, showing that for a multicast network, the minimum field size that satisfies the invertibility conditions in ${\cal G}_{ud}$ is at most as large as the minimum field size for ${\cal G}_{inst}.$ 

It is not difficult to observe that in some of the usual examples in network coding literature such as the butterfly network and combination networks, the feasibility of a given network code is preserved between the unit-delay network and the corresponding instantaneous network, because the topology of these networks prevents the mixing of information symbols from different generations at the intermediate nodes. In the forthcoming subsection, we formalize such a topological constraint for networks with general demands and thereby obtain sufficient conditions on the equivalence of network coding solutions between ${\cal G}_{ud}$ and ${\cal G}_{inst}$ for an acyclic network ${\cal G}$ with given demands.
\subsection{Equivalence of minimum field size problem between ${\cal G}_{inst}$ and ${\cal G}_{ud}$}
\label{subsec4b}
The following proposition gives a class of networks for which the minimum field size is equal for both ${\cal G}_{inst}$ and ${\cal G}_{ud},$ by demonstrating a sufficient condition under which certain network coding solutions remain feasible for both ${\cal G}_{inst}$ and ${\cal G}_{ud}$. We define for a node $v\in{\cal V}\backslash{\cal S},$ a set $Q(v)$ which consists of all possible paths (a path being a sequence of edges following an ancestral order) from the source nodes to $v$ such that any two paths differ by at least one edge. We also define for a node $v\in {\cal V}\backslash{\cal S},$ a $|Q(v)|$-length \textit{depth vector} $\boldsymbol{d}(v),$ each component (in $\mathbb{Z}^+$) of which indicates the total delay incurred in the corresponding path of $Q(v)$ from some source $s$ to node $v.$
\begin{proposition}
\label{prop3}
Let ${\cal G}_{ud}({\cal V},{\cal E},{\cal S},{\cal T},{\cal C})$ be an acyclic, unit-delay network with a given set of sink demands and ${\cal G}_{inst}({\cal V},{\cal E},{\cal S},{\cal T},{\cal C})$ be the corresponding instantaneous network. Suppose the topology of ${\cal G}_{ud}$ is such that for any $v\in {\cal V}\backslash{\cal S},$ the components of the depth vector $\boldsymbol{d}(v)$ are all equal. Let $\cal U$ be the set of all feasible solutions for ${\cal G}_{ud}({\cal V},{\cal E},{\cal S},{\cal T},{\cal C})$ such that the sources combine information symbols without using memory, i.e. the symbols only from the current generation, and ${\cal U}_q$ be the subset of $\cal U$ with solutions from the field $\mathbb{F}_q.$  Then the following statements are true. 
\begin{enumerate}
\item[(A)] Any solution from $\cal U$ for ${\cal G}_{ud}$ is also a feasible solution for ${\cal G}_{inst}.$
\item[(B)] Any feasible solution for ${\cal G}_{inst}$ is a feasible solution for ${\cal G}_{ud}.$
\item[(C)] If $q_{min}$ is the minimum field size for which a feasible network code exists for ${\cal G}_{ud}$ and the subset ${\cal U}_{q_{min}}$ of $\cal U$ is non-empty, then $q_{min}$ is the minimum field size required for a feasible solution for ${\cal G}_{inst}$ too.
\end{enumerate}
\end{proposition}
\textit{Proof:} See Appendix \ref{prop3proof}.

Proposition \ref{prop3} formalizes the easily observed sufficient conditions for the same network codes to be solutions for both ${\cal G}_{inst}$ and ${\cal G}_{ud}.$ However, deriving necessary conditions for the same seems difficult. Proposition \ref{prop3} also leads to the following obvious corollary.
\begin{corollary}
\label{cor3}
Let ${\cal G}_{ud}({\cal V},{\cal E},{\cal S},{\cal T},{\cal C})$ be an acyclic, unit-delay network with a given set of sink demands and ${\cal G}_{inst}({\cal V},{\cal E},{\cal S},{\cal T},{\cal C})$ be the corresponding instantaneous network. Suppose the topology of ${\cal G}_{ud}$ is such that for any $v\in {\cal V}\backslash{\cal S},$ the components of the depth vector $\boldsymbol{d}(v)$ are all equal. If ${\cal G}_{ud}$ has no feasible solutions over some particular field $\mathbb{F}_q,$ then neither does ${\cal G}_{inst}.$
\end{corollary}
\subsection{Reduction of minimum field size in ${\cal G}_{ud}$ - Review of Linear Information Flow Algorithm}
\label{lifalg}
Proposition \ref{prop3} illustrates some network conditions which lead to the equivalence between finding the minimum required field size for a given set of demands on ${\cal G}_{ud}$ and ${\cal G}_{inst}$ of a given acyclic network ${\cal G}$. Example \ref{exm1} illustrated a situation where the disparity in the delays of the symbols arriving at a node prevented the possibility of obtaining a feasible network code. However, such delay disparity can also be  useful. In particular, because of this delay disparity, there exist networks in which the feasible network codes exist over a smaller field for ${\cal G}_{ud}$ compared to ${\cal G}_{inst}.$ Towards understanding how such situations can arise, we discuss a couple of examples.  
\begin{example}
\label{exm2}
Consider the network ${\cal G}$ shown in Fig. \ref{fig2}. The source $s$ has two sequences $x_1(z)$ and $x_2(z)$ to be transmitted to the six sinks $t_i:1\leq i \leq 6.$ This network is clearly a cascaded version of the usual butterfly network and the $\left(\begin{array}{c} 4 \\ 2 \end{array}\right)$ network. As in the case of the $\left(\begin{array}{c} 4 \\ 2 \end{array}\right)$ network, a feasible network coding solution for this network (either in ${\cal G}_{inst}$ or in ${\cal G}_{ud}$) implies that any two of the four global encoding vectors on the edges $e_i: 1 \leq i \leq 4$ should be linearly independent. Therefore, for ${\cal G}_{inst},$ a minimum field size of $3$ is required to construct a feasible network code. 

However, for ${\cal G}_{ud},$ a binary field is sufficient. Consider the usual network code over $\mathbb{F}_2$ in the butterfly subnetwork of the given network, where the global encoding vectors at the node $v_2$ are $\left(\begin{array}{c}z^2 \\0 \end{array}\right)$ and $\left(\begin{array}{c}z^4 \\z^4 \end{array}\right),$ while those at node $v_3$ are $\left(\begin{array}{c}0 \\z^2 \end{array}\right)$ and $\left(\begin{array}{c}z^4 \\z^4 \end{array}\right).$ Then the vectors $\left(\begin{array}{c}z^3 \\0 \end{array}\right),$ $\left(\begin{array}{c}z^5 \\z^5 \end{array}\right),$ $\left(\begin{array}{c}0 \\z^3 \end{array}\right)$ and $\left(\begin{array}{c}z^5 \\z^5+z^3 \end{array}\right)$ can be chosen as global encoding vectors for the edges $e_i: 1 \leq i \leq 4$ respectively, which render the network code feasible for ${\cal G}_{ud}.$ Therefore, even in a multicast situation, the minimum field size requirement of the unit-delay network can be smaller than that of the corresponding instantaneous network. Note that such a situation is made possible because of the difference in the delays between the incoming symbols at the two edges of $v_3.$ 
\begin{figure}[htbp]
\centering
\includegraphics[totalheight=3in,width=3in]{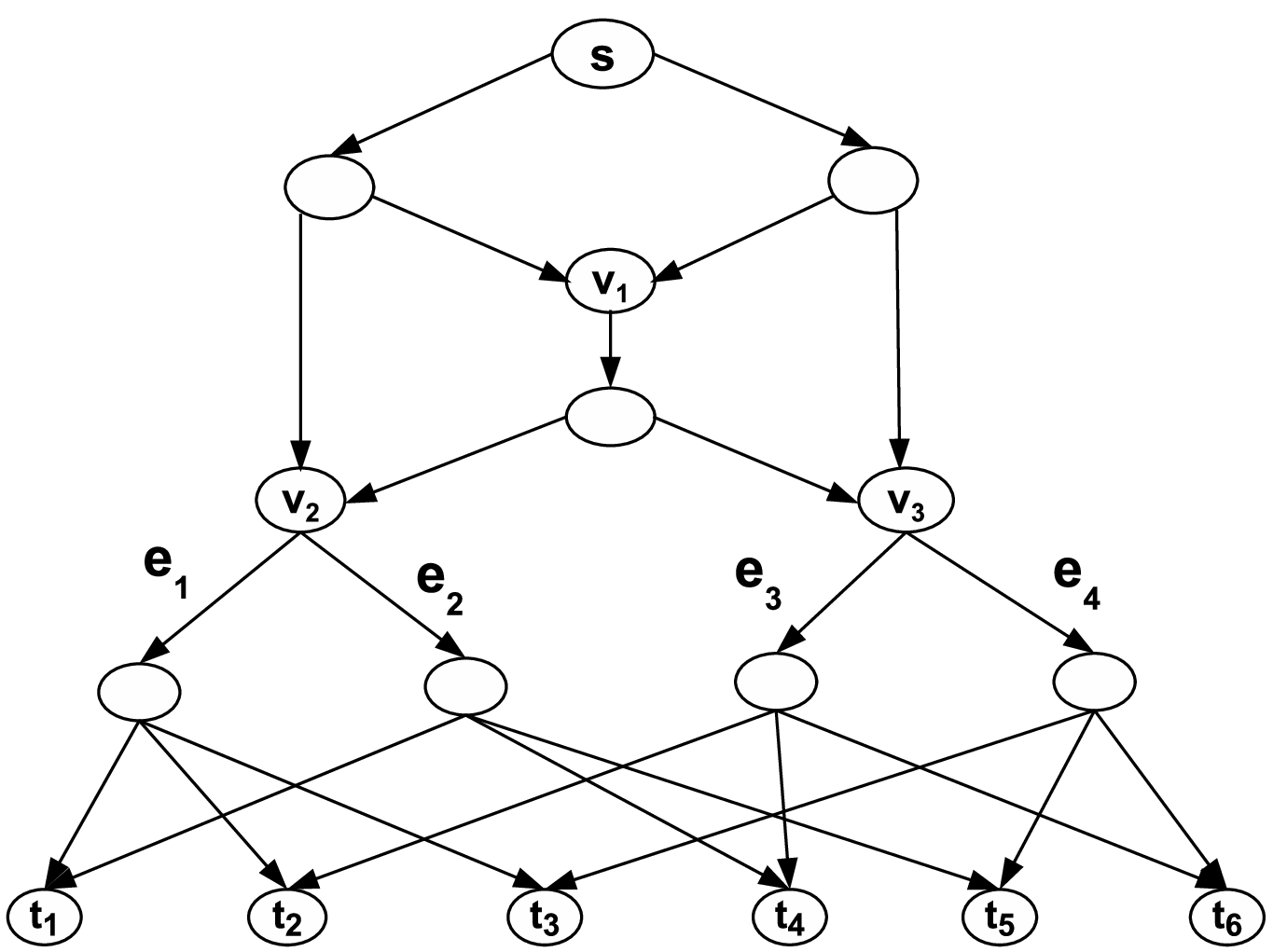}
\label{fig2}
\caption{A multicast network in which ${\cal G}_{ud}$ requires smaller field size than ${\cal G}$}	
\end{figure}
\end{example}
\begin{example}
\label{exm3}
Consider the network $\cal G$ shown in Fig. \ref{fig4}. For $1\leq j \leq 3,$ each source $s_j$ has an information sequence $x_j(z).$ This network has non-multicast demands, with sinks $t_i:1\leq i \leq 3$ requiring all three information sequences, while sink $t_4$ requires $\left\{x_1(z),x_3(z)\right\}$ and $t_5$ demands $\left\{x_2(z),x_3(z)\right\}.$ We show that no feasible network code exists for ${\cal G}_{inst}$ over $\mathbb{F}_2,$ while such a code exists for ${\cal G}_{ud}.$

We now argue that we cannot obtain a feasible network code for ${\cal G}_{inst}$ over $\mathbb{F}_2.$ The sinks $t_i: 1\leq i \leq 5$ have direct paths from the source(s) $\left\{s_1,s_2\right\}, \left\{s_1,s_3\right\}, \left\{s_2,s_3\right\},\left\{s_3\right\}$ and  $\left\{s_3\right\}$ respectively. As the sinks $t_i:1 \leq i \leq 3$ require information sequences from all three sources, the edge $e_1$ should carry a coded version of all three information sequences for the network code to be feasible at sinks $t_i: 1 \leq i \leq 3.$ Thus, over $\mathbb{F}_2,$  the global encoding vector for edge $e_1$ should be $\boldsymbol{a}=\left(1~~1~~1\right)^T.$ As sink $t_5$ has a direct path from source $s_3,$ the edge $e_2$ should carry a linear combination of both $x_2(z)$ and $x_3(z).$ Thus the global encoding vector of edge $e_2$ over $\mathbb{F}_2$ must be  $\boldsymbol{b}=\left(0~~1~~1\right)^T.$ Now, as sink $t_4$ has a direct path from source $s_3,$ the edge $e_3$ should carry a linear combination of both $x_1(z)$ and $x_3(z),$ i.e., $\left(1~~0~~1\right)^T.$ However, the global encoding vectors of the incoming edges at node $v_1$ are $\boldsymbol{a}$ and $\boldsymbol{b},$ using which the vector $\left(1~~0~~1\right)^T$ cannot be obtained. Thus no feasible network code can be found for ${\cal G}_{inst}$ over $\mathbb{F}_2.$ A feasible network code can be found for this network over any field with size $q\geq 3.$ 

Now we prove by argument that there exists a code for ${\cal G}_{ud}.$ Because $e_1$ should carry a linear combination of all three information sequences, let its global encoding vector be $\boldsymbol{a}(z)=\left(z^2~~z^2~~z^2\right)^T,$ after accounting for the delays incurred in the transmission. Accounting for the disparity in the delays at the node $v_2,$ let the global encoding vector of edge $e_2$ be $\boldsymbol{b}(z)=\left(0~~z^2~~z^3\right)^T.$ Thus the global encoding vectors of the incoming edges at node $v_1$ are $z\boldsymbol{a}(z)$ and $z\boldsymbol{b}(z).$ Node $v_1$ can then simply send a sum of two incoming symbols on edge $e_3,$ in which case the global encoding vector of edge $e_3$ is $\left(z^4~~0~~z^4+z^5\right)^T.$ Accounting for all the direct paths in the network, it can be seen that all sink demands are satisfied, i.e., the invertibility conditions hold over $\mathbb{F}_2(z)$ and so do the zero-interference conditions. Thus there is a feasible network code for ${\cal G}_{ud}$ over $\mathbb{F}_2.$ As in the previous example, the delay disparity at node $v_2$ is what makes this possible.
\begin{figure}[htbp]
\centering
\includegraphics[totalheight=2.8in]{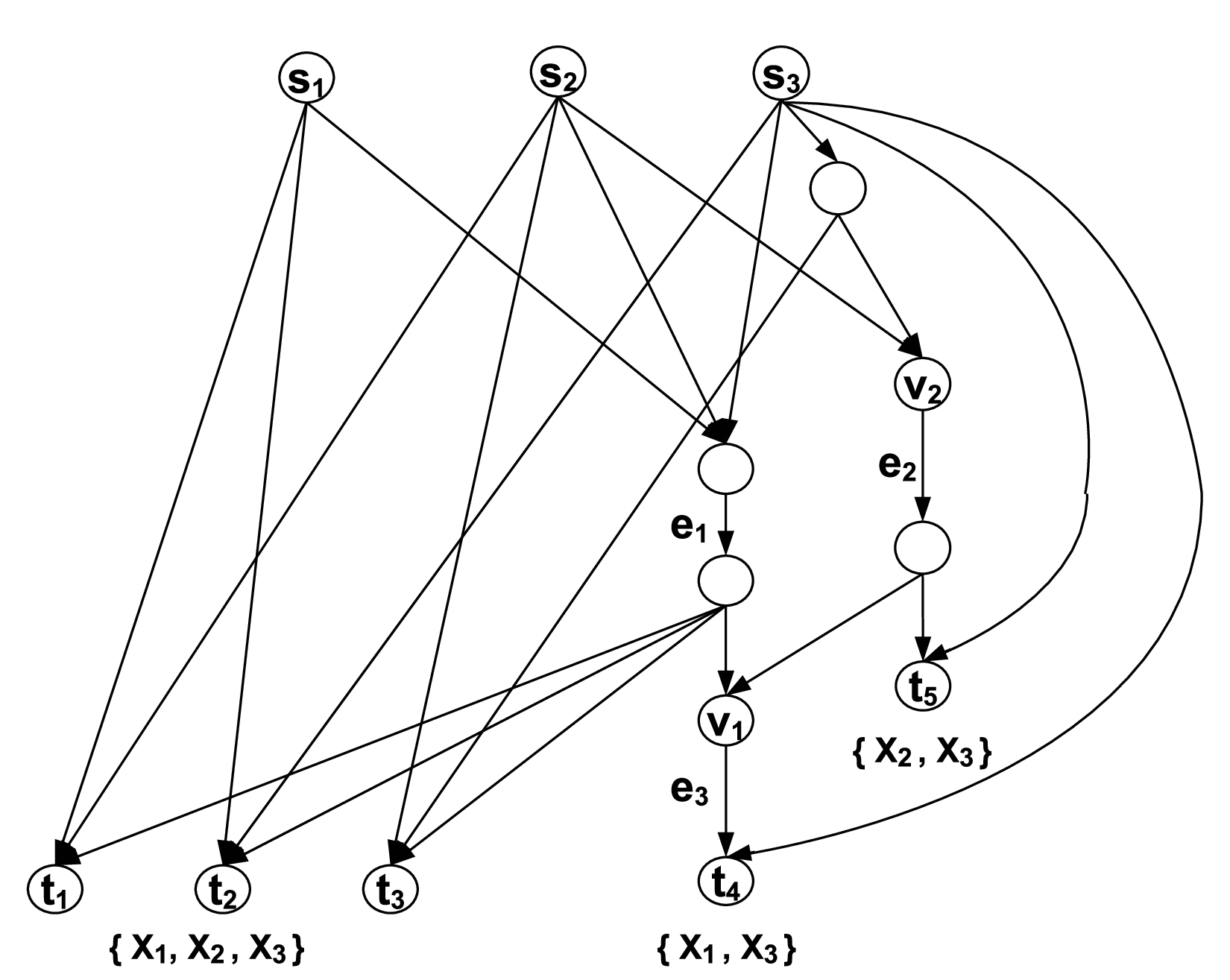}
\label{fig4}	
\caption{A non-multicast network in which ${\cal G}_{ud}$ requires smaller field size than ${\cal G}$}
\end{figure}
\end{example}

In the remainder of this section, we show that there exist several occasions in which a binary field is sufficient for constructing feasible network codes for unit delay networks, irrespective of the field size required for their instantaneous counterparts. We concentrate only on the single-source multicast case. As the results in this section are all centered on the deterministic version of the \textit{Linear Information Flow} (LIF) algorithm in \cite{JSCEEJT}, we briefly discuss the terminology and lemmas related to the LIF algorithm before proving our results.

The steps of the LIF algorithm for constructing a multicast network code in a single source acyclic instantaneous network ${\cal G}_{inst}$ with $|{\cal T}|$ sinks are as follows. 
%
\begin{enumerate}
\item Identify the $h_s$ edge-disjoint paths from the source to the sinks, where $h_s$ is the number of information symbols at the source. Note that $h_s$ can be at most equal to the minimum of the mincuts between the source and each sink and not more, otherwise the  multicast problem is infeasible. Let ${\cal G}^*({\cal V}^*,{\cal E}^*)$ be the subnetwork of ${\cal G}$ consisting of the nodes and edges on these edge-disjoint paths alone. The rest of the algorithm works only with ${\cal G}^*,$ as a feasible network code for ${\cal G}^*$ can be converted to a feasible network code for ${\cal G}$ by simply assigning zeros for any other local encoding coefficients.
\item For a sink $t \in {\cal T},$ let $f^t$ denote the set of $h_s$ edge-disjoint paths from the source to $t.$ For each sink $t\in \cal T,$ the algorithm maintains for each sink a set $C_t,$ which consists of the $h_s$ most recently processed edges (one from every path	in $f^t$) and a $h_s \times h_s$ matrix $B_t,$ which has the global encoding vectors of the edges in $C_t.$ The set $C_t$ is initialized with the set $\left\{e_i:1\leq i \leq h_s\right\}$ (the $h_s$ imaginary edges at the source), while the columns of the matrix $B_t$ is initialized with the $h_s$-length global encoding vectors $\left\{\boldsymbol{b}(e_i):1\leq i \leq h_s\right\}$ respectively.  For every sink $t,$ the algorithm maintains the full-rank property of $B_t$ by an appropriate choice of the local encoding coefficients at $tail(e),$ $e\in C_t$ being the most recently processed edge  as $C_t$ is incremented in some ancestral ordering. The algorithm also maintains another $h_s \times h_s$ matrix $A_t$ which has the inverse vectors of $B_t$ at every step of the algorithm. 
\item 
For an edge $e \in {\cal E}^*$, let $f^t_{\leftarrow}(e)$ denote the predecessor edge of $e$ on some flow path from the source $s$ to sink $t.$ Note that there can be at most one flowpath from $s$ to any sink through $e.$ After processing any edge $e$, the LIF algorithm updates the set $C_t$ and the matrices $B_t$ and $A_t.$ The updated values for $C_t, B_t$ and $A_t$ are denoted by $\tilde{C}_t, \tilde{B}_t$ and $\tilde{A}_t$ and are obtained as follows.
\begin{enumerate}
\item
$
\tilde{C}_t = C_t\backslash\left\{f^{t}_{\leftarrow}(e)\right\}\cup\left\{e\right\}.
$
\item 
$
\tilde{B}_t = \tilde{B}_t
\backslash\left\{\boldsymbol{b}(f^{t}_{\leftarrow}(e))\right\}\cup\left\{\boldsymbol{b}(e)\right\}.
$
\item
$
\boldsymbol{\tilde{a}_t}(e)=\left(\boldsymbol{b}(e).\boldsymbol{a_t}(f^{t}_{\leftarrow}(e))\right)^{-1}\boldsymbol{a_t}(f^{t}_{\leftarrow}(e)).
$
\item 
For all $c \in C_t\backslash\left\{f^{t}_{\leftarrow}(e)\right\},$
\begin{align}
\label{eqn15}
\boldsymbol{\tilde{a}_t}(c)=\boldsymbol{a_t}(c)-\left(\boldsymbol{b}(e).\boldsymbol{a_t}(c)\right)\boldsymbol{\tilde{a}_t}(c).
\end{align}
\end{enumerate}
\item Ultimately the algorithm ends with the matrix $B_t$ equal to the network transfer matrix $M_t$ of the sink $t,$ whose full-rank property is guaranteed by the algorithm, therefore rendering a feasible network code. 

\end{enumerate}

Let $T(e)$ be the number of sinks which have a flow path through $e,$ and let $P(e)=\left\{f^t_{\leftarrow}(e): t \in T(e) \right\}$ be the set of all predecessor edges of $e.$ The only non-zero local encoding coefficients to be chosen for the edge $e$ by the algorithm are $m_{e,p},~p\in P(e).$ At every step of the algorithm where the global encoding vector of the next edge (according to a chosen ancestral order) $e$ is selected, the set $B_t$ has to be kept linearly independent, i.e., the choice of $m_{e,p}$ should be such that $B_t\backslash\left\{\boldsymbol{b}(f^t_{\leftarrow}(e))\right\}\cup\left\{\boldsymbol{b}(e)\right\}$ is linearly independent. The following lemma proved in \cite{JSCEEJT} gives a fast way to test this linear independence based on the dot product in $\mathbb{F}_q.$ In the following lemma, we have $\delta_{a,b}=1,$ if $a=b.$
\begin{lemma}
\label{lemma1}
Consider a basis $B$ of $\mathbb{F}_q^{h_s}$ and vectors $\boldsymbol{b} \in B,$ $\boldsymbol{a}\in \mathbb{F}_q^{h_s}$ such that $\forall \boldsymbol{b'}\in B,$ we have $\boldsymbol{b'}.\boldsymbol{a} = \delta_{\boldsymbol{b},\boldsymbol{b'}}.$ Then, any vector $\boldsymbol{x} \in \mathbb{F}_q^{h_s}$ is linearly dependent on $B\backslash\left\{\boldsymbol{b}\right\}$ if and only if $\boldsymbol{x}.\boldsymbol{a}=0.$
\end{lemma}

Given the full-rank matrix $B_t$ of global encoding vectors $\left\{\boldsymbol{b}(e) \in \mathbb{F}_q^{h_s}: e \in C_t \right\},$ we will denote the corresponding columns of the inverse matrix $A_t$ of $B_t,$ as $\left\{\boldsymbol{a_t}(e) \in \mathbb{F}_q^{h_s}: e \in C_t \right\}.$ Then the linear independence condition to be checked in the LIF algorithm takes the following form due to Lemma \ref{lemma1}: 
\begin{equation}
\label{eqn17}
\forall t\in T: \forall e,e'\in C_t:\boldsymbol{b}(e).\boldsymbol{a_t}(e')= \delta_{e,e'}.
\end{equation}

The following lemma gives the sufficient field size for the construction of a feasible network code for multicast on a single source acyclic network. 
\begin{lemma}\cite{JSCEEJT}
\label{lemma2}
Let $q \geq n.$ Consider pairs $(\boldsymbol{x_i},\boldsymbol{y_i}) \in \mathbb{F}_q^{h_s}\times \mathbb{F}_q^{h_s}$ with $\boldsymbol{x_i}.\boldsymbol{y_i} \neq 0$ for $i \leq i \leq n.$ There exists a linear combination $\boldsymbol{u}$ of $\boldsymbol{x_1},\boldsymbol{x_2},...,\boldsymbol{x_n}$ such that $\boldsymbol{u}.\boldsymbol{y_i} \neq 0$ for $1 \leq i \leq n.$ 
\end{lemma}

\textit{Outline of proof:}
We provide only an outline here in order that we might use similar proof ideas later in Subsection \ref{delayandcode}. For the complete proof, the reader is referred to \cite{JSCEEJT}. The proof involves the iterative construction of vectors  $\boldsymbol{u_1},\boldsymbol{u_2},...,\boldsymbol{u_n}$ (each $\boldsymbol{u_i} \in \mathbb{F}_q^{h_s}$) such that for any $i,$ $\boldsymbol{u_i}$ is some linear combination of the vectors $\left\{\boldsymbol{x_k}: 1\leq k \leq i\right\}$ and for any $j$ such that $1\leq j \leq i,$ $\boldsymbol{u_i}.\boldsymbol{y_j} \neq 0.$ As long as the field size is more than $n,$ the vectors $\boldsymbol{u_i}$ can always be found, with the final vector $\boldsymbol{u_n}$ being the desired $\boldsymbol{u}.$ \hspace*{\fill}{$\blacksquare$}
   

With $e$ being the edge under consideration, let  
\begin{equation*}
\left\{(\boldsymbol{x_i},\boldsymbol{y_i}):i \leq i \leq n\right\}=\left\{(\boldsymbol{b}(f^t_{\leftarrow}(e)),\boldsymbol{a_t}(f^t_{\leftarrow}(e))):t\in T(e) \right\}
\end{equation*}
in Lemma \ref{lemma2}. Then the vector $\boldsymbol{u}$ found using Lemma \ref{lemma2} satisfies (by invoking Lemma \ref{lemma1}) the requirements for the global encoding vector $\boldsymbol{b}(e)$ of edge $e,$ i.e., (\ref{eqn17}). The particular linear combination of the vectors $\left\{\boldsymbol{b}(f^t_{\leftarrow}(e)): e \in P(e)\right\}$ used to obtain $\boldsymbol{u}$ gives the local encoding coefficients at $tail(e),$ i.e., $\left\{m_{e,p}: p\in P(e)\right\}.$ By Lemma \ref{lemma2}, a field size $q$ such that
\begin{equation}
\label{eqn9}
q > \max_{e\in{\cal E}}T(e),
\end{equation}
$T(e)$ being the number of sinks which have flow paths through $e,$ is always sufficient for constructing a multicast network code for ${\cal G}^*_{inst}$ according to the LIF algorithm. Therefore, for constructing a multicast network code in any single source acyclic network with $|{\cal T}|$ sinks,  $q > |{\cal T}|$ is sufficient.

Note that although the LIF designs a feasible network code for ${\cal G}_{inst},$ the extension of the LIF algorithm (and the associated lemmas) to ${\cal G}_{ud}$ is straightforward. While the local encoding coefficients (picked according to Lemma \ref{lemma2}) continue to be over $\mathbb{F}_q,$ the matrices $A_t$ and $B_t$ are over $\mathbb{F}_q(z)$ for ${\cal G}_{ud}$ according to (\ref{eqn4}). Therefore, the dot product involved in the Lemma \ref{lemma1} and Lemma \ref{lemma2} are the standard dot product in $\mathbb{F}_q(z),$ and  full-rank property of $B_t$ is checked over $\mathbb{F}_q(z).$ By Corollary \ref{cor1}, (\ref{eqn9}) holds for multicast network code construction for ${\cal G}^*_{ud}$ too.
\subsection{Delay-and-code: A technique for single-source multicast on acyclic networks}
\label{delayandcode}
As discussed in Subsection \ref{lifalg}, the LIF algorithm uses Lemma \ref{lemma2} together with Lemma \ref{lemma1} for constructing a multicast network code in a given instantaneous  network ${\cal G}_{inst},$ or the corresponding unit-delay network ${\cal G}_{ud}.$ Based on the LIF algorithm, we now present another network coding scheme called a \textit{delay-and-code} scheme, which reduces the complexity of encoding at the intermediate nodes at the cost of potentially increased complexity of decoding at the sink nodes. The finite field under consideration is always $\mathbb{F}_2.$ As a theoretical by-product of this scheme, we show that there exist networks for which the binary field is sufficient for constructing a multicast network code in ${\cal G}_{ud},$ irrespective of the field size requirement in ${\cal G}_{inst}.$

We assume that each node is equipped with memory elements and a linear combination of the stored symbols is then transmitted on the outgoing edges. Abusing the definition of the delay parameter $z,$ we also denote a memory element by $z.$ We however do not allow all possible $\mathbb{F}_2(z)$-linear combinations of the incoming symbols that is possible using the memory elements available at the node under consideration. In other words, the input-output relationship of the edge $e$ given by (\ref{eqn4}) is restricted to be of the form
\begin{equation}
\label{eqn10}
y_e(z)=\sum_{p\in\Gamma_I(tail(e))}m_{e,p}z^{a_{e,p}}y_p(z),
\end{equation}
where $m_{e,p} \in \mathbb{F}_2,$ and $a_{e,p} \in \mathbb{Z}^{\geq 0},$ in general (for both ${\cal G}_{inst}$ and ${\cal G}_{ud}$). For ${\cal G}_{ud},$ $a_{e,p} \in \mathbb{Z}^+,$ to account for mandatory delay incurred in the transmission through edge $e.$

If a delay-and-code scheme on $\cal G$ is such that for any non-source node $v\in {\cal V}\cup{\cal T},$ and for any $p \in \Gamma_{I}(v),$ 
\begin{equation}
\label{eqn11}
a_{e,p} = a_p,~\forall e\in\Gamma_{O}(v)\text{ such that }m_{e,p}\neq 0,
\end{equation}
then we refer to the delay-and-code scheme as a \textit{uniform delay-and-code} scheme. Otherwise, we refer to it as a \textit{non-uniform delay-and-code} scheme. In other words, in the uniform delay-and-code scheme, an intermediate node is not allowed to code differently delayed versions of the symbols arriving from any particular edge. In the non-uniform case, this is permitted. Note that we consider only intermediate nodes in the network, i.e., the non-source non-sink nodes. The non-uniform delay-and-code technique was already mentioned in \cite{BaY2} for acyclic and cyclic networks. It was however only conjectured that a feasible multicast network code can be designed using the non-uniform delay-and-code scheme. In this work, we prove this conjecture for the case of acyclic networks.

Similar to the usual network coding formulation, such linear combinations also result in a network transfer matrix at each sink, which should be full-rank over $\mathbb{F}_2(z)$ for the network code to be feasible for that particular sink. The algorithm for constructing a delay-and-code scheme for any given acyclic single-source network with multicast demands follows that of the LIF algorithm, with the change that the local encoding coefficients are based on the formulation of (\ref{eqn10}). The following proposition shows that any acyclic network with multicast demands can be solved using the non-uniform delay-and-code scheme. 
\begin{proposition}
\label{prop4}
Let ${\cal G}({\cal V},{\cal E},s,{\cal T},{\cal C})$ be an acyclic single-source network with multicast demands with the mincut between $s$ and any $t\in\cal T$ being at least $h_s,$ the number of information sequences generated at $s.$ Then a feasible network code can be designed for ${\cal G}_{inst}$ (or ${\cal G}_{ud}$) using the non-uniform delay-and-code scheme provided the total number of memory elements present at each node for each incoming edge is at least $\left(|{\cal T}|-1\right).$ 
\end{proposition}
\textit{Proof:} See Appendix \ref{prop4proof}.

We now deal with the uniform delay-and-code scheme. We consider only the special case of those networks in which the paths from the source to each sink are not only edge-disjoint but also node-disjoint, i.e., the $h_s$ paths from the source to any sink do not have any common node except the source and that particular sink itself. The general case, where paths are not necessarily node-disjoint, is more difficult and might not be solvable because of the following reason. 

Consider a network with paths that are not node-disjoint. Because of the formulation specified by (\ref{eqn11}), for any given intermediate node $v,$ all the priorly processed outgoing edges of $v$ should be considered when processing any particular $e' \in \Gamma_O(v).$ Let $e,e' \in \Gamma_O(v)$ and suppose $T(e)\cap T(e') \neq \Phi.$ Suppose the global encoding vector of $e$ has been decided before $e'.$ Once edge $e'$ has been processed ($\boldsymbol{b}(e')$ has been decided),  note that the elements of the set $S(e),$ which was used to determine $\boldsymbol{b}(e),$ would have been updated according to (\ref{eqn15}). Thereby $\boldsymbol{b}(e)$ might no longer satisfy the required properties of maintaining the ranks of the $B_t$ matrices of some $t\in T(e)\cap T(e').$ Now if a new global encoding vector $\boldsymbol{b}(e)$ was chosen for edge $e,$ then the set $S(e')$ might change, and $\boldsymbol{b}(e')$ might no longer be a valid global encoding vector for $e'.$ Because of such a see-saw effect, it might not be possible to design a feasible network code using the uniform delay-and-code scheme. 

Now suppose the $h_s$ edge-disjoint paths to each sink from $s$ are also node-disjoint, i.e., at any intermediate node in the network, there exists at most one incoming-outgoing edge pair which lies on any path from the source to any particular sink. Therefore, for any $e \in \Gamma_O(v)$ being the currently processed edge, any $c \in C_t\backslash\left\{f^{t}_{\leftarrow}(e)\right\}$ for any $t \in T(e)$ is such that $c\notin \Gamma_I(v).$ In other words, there are no $e,e' \in \Gamma_O(v)$ such that $T(e)\cap T(e') \neq \Phi.$ Thus, fixing the global encoding vector for any $e'\in\Gamma_O(v)$ does not affect the sets $S(e)$ for any other edge $e\in\Gamma_O(v).$ For this reason, we focus on networks with node-disjoint paths for the uniform delay-and-code case.
\begin{proposition}
\label{prop5}
Let ${\cal G}({\cal V},{\cal E},s,{\cal T},{\cal C})$ be an acyclic single-source network with multicast demands with at least $h_s$ node-disjoint paths between $s$ and any $t\in\cal T,$ $h_s$ being the number of information sequences generated at $s.$ Let $\delta:=\max_{v\in \cal V} \delta_O(v).$ Then a feasible network code can be designed for ${\cal G}_{inst}$ (or ${\cal G}_{ud}$) using the uniform delay-and-code scheme provided the total number of memory elements present at each node for each incoming edge is at least $\delta\left(|{\cal T}|-1\right).$ 
\end{proposition}
\textit{Proof:} See Appendix \ref{prop5proof}.

The following corollary to Proposition \ref{prop5} shows that there exist several unit-delay networks for which a binary field is sufficient for constructing a feasible network code, irrespective of the field size required for their instantaneous counterparts.
\begin{corollary}
\label{cor4}
There exist acyclic networks for which a feasible binary network code exists for the unit-delay networks as a result of differently delayed information available at the coding nodes (where paths to different sinks intersect) in the network, irrespective of the minimum field size required to design a feasible network code for the corresponding instantaneous networks. In particular, given a single-source acyclic network $\cal G$ with multicast demands and with at least $h_s$ node-disjoint paths ($h_s$ being the number of information symbols at source) from the source to each sink, it is always possible to construct a modified network $\tilde{{\cal G}}$ such that $\tilde{{\cal G}}_{inst}$ has the same minimum field size requirement as ${\cal G}_{inst},$ but a binary field size would suffice to obtain a feasible network code  for $\tilde{{\cal G}}_{ud}.$
\end{corollary}
\textit{Proof:} See Appendix \ref{cor4proof}.

Although the delay-and-code scheme can be used to construct feasible network codes in the multicast situation, for a network with more general demands, it might not prove to be useful. We now present an example where it is not possible to design a network code using the delay-and-code scheme. 
\begin{example}
\begin{figure}[htbp]
\centering
\includegraphics[width=2.5in]{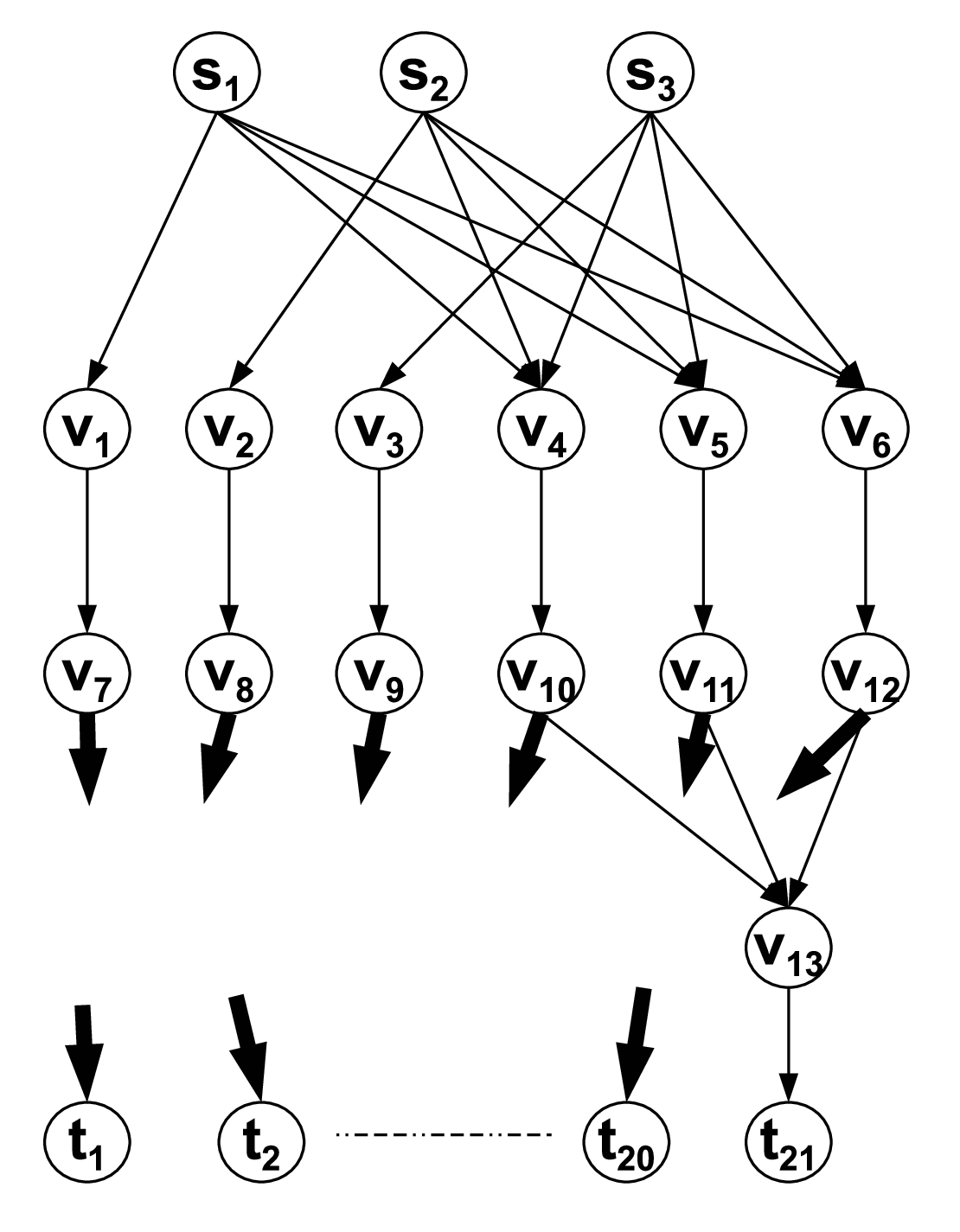}
\caption{A network for which the delay-and-code scheme cannot be used to construct a feasible network code}	
\label{fig5}	
\end{figure}
Consider the network $\cal G$ shown in Fig. \ref{fig5}, with sources $\left\{s_i:1 \leq i \leq 3\right\}$ and sinks $\left\{t_i: 1 \leq i \leq 21\right\}.$ The source $s_i$ generates the information sequence $x_i(z)$. The subnetwork of $\cal G$ consisting of all nodes and edges of $\cal G$ except the nodes $v_{13}$ and sink $t_{21}$ is derived from the $\left(\begin{array}{c} 6 \\ 3 \end{array}\right)$ combination network, i.e., for every possible three-combination of the nodes $\left\{v_i:7\leq i\leq 12\right\},$ there exists a sink to which there is precisely one edge incoming from the three nodes, each of which demands all three of the information sequences. There are therefore $\left(\begin{array}{c} 6 \\ 3 \end{array}\right)=20$ such sinks, and the bolded arrows indicate the ten outgoing edges from each node $v_i:7\leq i\leq 12,$ and the three incoming edges from $\left\{v_i:7\leq i\leq 12\right\}$ to each sink $t_i:1\leq i \leq 20.$ Each of these $20$ sinks demand all three information sequences. The additional sink $t_{21}$ demands the information sequence $x_1(z).$ Note that there exists a solution to this network (for both ${\cal G}_{inst}$ and ${\cal G}_{ud}$) if the field size $q\geq 4.$

We now attempt to obtain a delay-and-code scheme on this network (in either ${\cal G}_{inst}$ or ${\cal G}_{ud}$). Because each node $v_i$ is only connected to the source $s_i$ for $1\leq i \leq 3,$ each of the global encoding vectors of the outgoing edges from the nodes $\left\{v_i:1\leq i \leq 3\right\}$ has one component of the form $z^{a},$ $a$ being some non-negative integer, with the other two components being zero. Also, the three global encoding vectors of the outgoing edges from the nodes $\left\{v_i:4\leq i \leq 6\right\}$ have to be linearly independent and of the form $\left(z^a~~z^b~~z^c\right)^T,$ because each of these three vectors have to be linearly independent with any two of the three global encoding vectors of the outgoing edges from $\left\{v_i:1\leq i \leq 3\right\}.$ For $4\leq i \leq 6,$ let the global encoding vector of node $v_i$ be $\boldsymbol{f}_i=\left(z^{a_i}~~z^{b_i}~~z^{c_i}\right)^T,$ for some non-negative integers $a_i,b_i$ and $c_i.$ 

To satisfy the requirements for sink $t_{21},$ a delay-and-code based linear combination of the  vectors $\boldsymbol{f}_i:4\leq i\leq 6$ should generate a vector of the form $\left(z^d~~0~~0\right)^T,$ for some non-negative integer $d.$ In other words, for some non-negative integers $\tilde{a},\tilde{b}$ and $\tilde{c},$ and for some $m_1, m_2, m_3 \in \mathbb{F}_2,$ we want
\begin{equation*}
\left(
\begin{array}{ccc}
z^{a_4}& z^{a_5}&z^{a_6}\\
z^{b_4}&z^{b_5}&z^{b_6}\\
z^{c_4}&z^{c_5}&z^{c_6}
\end{array}
\right)
\left(
\begin{array}{c}
m_1z^{\tilde{a}}\\
m_2z^{\tilde{b}}\\
m_3z^{\tilde{c}}
\end{array}
\right) =
\left(
\begin{array}{c}
z^d\\
0\\
0
\end{array}
\right).
\end{equation*}

Note that $m_1, m_2$ and $m_3$ cannot all be $0.$  They cannot  all be $1,$ as it is not possible to find non-negative integers $\tilde{a},\tilde{b}$ and $\tilde{c}$ such that $z^{\tilde{a}+b_4}+z^{\tilde{b}+b_5}+z^{\tilde{c}+b_6}=0,$ or such that $z^{\tilde{a}+c_4}+z^{\tilde{b}+c_5}+z^{\tilde{c}+c_6}=0.$ 

Now, suppose two of $m_1, m_2$ and $m_3$ are non-zero, then this means that the global encoding vector of the outgoing edge from node $v_1$ lies in the space spanned by two of the global encoding vectors of the outgoing edges from $\left\{v_i:4 \leq i \leq 6 \right\}.$ But this contradicts our original choice of these global encoding vectors, according to which any two are linearly independent with the global encoding vector of the outgoing edge from node $v_1.$ Therefore, a delay-and-code based scheme cannot satisfy the requirements of all sinks in this network, in either ${\cal G}_{inst}$ or ${\cal G}_{ud}.$ 
\end{example}
\section{Concluding remarks}
\label{sec5}
We have discussed the effects of using the delay inherent in the network in problems related to network code existence and designs. The delay-and-code algorithms presented in this paper enable low-complexity encoding at the intermediate nodes at the cost of using large memories for decoding at the sinks. A simple upper bound for the maximum number of memory elements required at any sink to decode the information sequences which are encoded using a delay-and-code scheme can be obtained without much difficulty. Similar algorithms can be found in \cite{BaY2,HLu,KeK}. Also, while the equivalence between memory elements and delays might not in practice make sense as the actual value of the delay incurred in the two might not be equal, the parameter $z$ used can be equivalently used to express both and therefore Corollary \ref{cor4} still holds. 

The results obtained in this work indicate that using delays in the network might be beneficial in certain situations, while being not useful in others. In any case, the delays in the network cannot be ignored for analyzing any network coding problem. Subsequent work might include the analysis of random network coding in unit-delay networks and the study of cyclic networks in a similar manner.

\appendices
\section{Proof of Proposition \ref{prop1}}
\label{prop1proof}
\begin{IEEEproof}
Let $M'_t$ be the $h_t \times h_t$ submatrix of the network transfer matrix of any particular sink node $t \in \cal T$ in ${\cal G}_{inst}$, involving the $h_t$ information symbols to be inverted. Let $M'_t(z)$ be the corresponding matrix of the same sink $t$ in ${\cal G}_{ud}.$ We first note that the matrix $M'_t$ can be obtained from $M'_t(z)$ by substituting $z=z^0=1$, i.e., 
\[
M'_t = M'_t(z)|_{z=1}.
\]
Given that $M'_t$ is of full rank over $\mathbb{F}_q$, we prove that $M'_t(z)$ is of full rank over $\mathbb{F}_q(z)$ by contradiction.

Suppose that $M'_t(z)$ was not of full rank over $\mathbb{F}_q(z)$, then we have 
\begin{equation}
\label{eqn1}
\sum_{i=1}^{h_t-1}\frac{a_i(z)}{b_i(z)}\boldsymbol{M'_i}(z) = \boldsymbol{M'_{h_t}}(z),
\end{equation}
where $\boldsymbol{M'_i}(z)$ is the $i^{th}$ column of $M'_t(z)$ and $a_i(z), b_i(z) \in \mathbb{F}_q[z]$ $\forall$ $i = 1,2,..,h_t-1$ are such that $b_i(z) \neq 0~\forall  i,$ and $a_i(z) \neq 0$  for at least one $i$, and $gcd(a_i(z), b_i(z))=1, ~\forall  i.$ We have the following two cases

\textit{Case 1}: $b_i(z)|_{z=1} \neq 0$ $\forall  i.$

Substituting $z = 1$ in (\ref{eqn1}), we have 
\begin{equation}
\label{eqn3}
\sum_{i=1}^{h_t-1}\frac{a_i}{b_i}\boldsymbol{M'_i} = \boldsymbol{M'_{h_t}},
\end{equation}
where $a_i=a_i(z)|_{z=1}, b_i=b_i(z)|_{z=1}$ and $\boldsymbol{M'_i}=\boldsymbol{M'_i}(z)|_{z=1}$ is the $i^{th}$ column of $M'_t.$ 

Clearly $\boldsymbol{M'_{h_t}} \neq \boldsymbol{0}$ since $M'_t$ is of full rank, and hence the left hand side of (\ref{eqn3}) cannot be zero. Therefore, some non-zero linear combination of the first ${h_t}-1$ columns of $M'_t$ is equal to its ${h_t}^{th}$ column, which contradicts the given statement that $M'_t$ is of full rank over $\mathbb{F}_q.$ Therefore, $M'_t(z)$ must be of full rank over $\mathbb{F}_q(z).$

\textit{Case 2}:  $b_i(z)|_{z=1} = 0$ for at least one $i.$

Let ${\cal I}' \subseteq \left\{1,2,...,{h_t}\right\}$ be such that $(z-1)^{p'}|b_i(z)$ for some positive integer $p'.$ Let $p$ be an integer such that 
\[
p = \max_{i \in {\cal I}'}{p'}.
\]
Now, from (\ref{eqn1}) we have
\begin{equation}
\label{eqn2}
\sum_{i=1}^{{h_t}-1}(z-1)^{p}\frac{a_i(z)}{b_i(z)}\boldsymbol{M'_i}(z) =(z-1)^{p} \boldsymbol{M'_{h_t}}(z).
\end{equation}
Let ${\cal I} \subseteq \left\{1,2,..,{h_t}\right\}$ be such that $(z-1)^{p}|b_i(z)$ $\forall$ $i \in \cal I.$ Then, we must have that $(z-1) \nmid a_i(z)$ $\forall$ $i \in \cal I,$ since $gcd(a_i(z),b_i(z))=1.$ Also, let $b_i'(z) = b_i(z)/(z-1)^{p} \in \mathbb{F}_q[z]$ $\forall$ $i \in \cal I.$ Then we have 
\[
\left((z-1)^{p} \frac{a_i(z)}{b_i(z)}\right)|_{z=1}=\left(\frac{a_i(z)}{b_i'(z)}\right)|_{z=1} = \frac{a_i}{b_i'} \in \mathbb{F}_q\backslash\left\{0\right\},
\]
where $b_i' = b_i'(z)|_{z=1}\in \mathbb{F}_q\backslash\left\{0\right\}$, since ${(z-1)}\nmid{b_i'(z).}$ Substituting $z=1$ in (\ref{eqn2}), we have 
\[
\sum_{i\in {\cal I}} \frac{a_i}{b_i'}\boldsymbol{M'_i} = \boldsymbol{0},
\]
i.e., a non-zero linear combination of the columns of $M'_t$ is equal to zero, which contradicts the full-rankness of $M'_t$, thus proving that $M'_t(z)$ has to be of full rank over $\mathbb{F}_q(z).$ As the choice of the sink $t$ was arbitrary, this completes the proof.
\end{IEEEproof}
\section{Proof of Proposition \ref{prop2}}
\label{prop2proof}
\begin{IEEEproof}
Let $\boldsymbol{m'}$ be a set of local encoding coefficients taking values from some field $\mathbb{F}_q$ which result in a feasible network code for ${\cal G}_{ud}$, satisfying the invertibility and zero-interference conditions, i.e., $f_1(\boldsymbol{m'},z)=f_2(\boldsymbol{m'},z)=...=f_K(\boldsymbol{m'},z)=0$ and the product of the determinants $g(\boldsymbol{m'},z)= \frac{g_n(\boldsymbol{m'},z)}{g_d(\boldsymbol{m'},z)} \neq 0,$ where $g_n(\boldsymbol{m'},z), g_n(\boldsymbol{m'},z) \in \mathbb{F}_q[z],$ the ring of polynomials in variable $z$ over $\mathbb{F}_q[z],$ are the numerator and denominator polynomials corresponding to $g.$

Note that, if we assign some appropriate value in $\mathbb{F}_q$ for the parameter $z$ in (\ref{eqn4}), we get a well-defined network code for ${\cal G}_{inst}$. In other words, if throughout the network, we let $z=z_{q}\in\mathbb{F}_q$ such that $\left(z-z_{q}\right)$ does not divide any numerator polynomial of $m_{e,p}(z)$ corresponding to any pair of edges $e$ and $p,$ then the unit-delay equation 
\begin{equation*}
y_e(z)=\sum_{p\in\Gamma_I(tail(e))}m'_{e,p}(z)|_{z=z_{q}}y_p(z) 
\end{equation*}
reduces to the instantaneous form, without the time index, as 
\begin{equation*}
y_e=\sum_{p\in\Gamma_I(tail(e))}m''_{e,p}y_p, 
\end{equation*}
where $m''_{e,p}=m'_{e,p}(z)|_{z=z_{q}},$ for all pairs of edges $e$ and $p$ in the network. The new set of local encoding coefficients, denoted by $\boldsymbol{m''}$ is a well-defined network code for ${\cal G}_{inst}$ (although this might not be feasible). Using this technique, we now show that given a feasible network code ($\boldsymbol{m'}$) for ${\cal G}_{ud}$ over some field $\mathbb{F}_q,$ we can obtain a feasible network code for ${\cal G}_{inst},$ over a possibly larger field $\mathbb{F}_{Q}.$

Given that $g(\boldsymbol{m'},z)= \frac{g_n(\boldsymbol{m'},z)}{g_d(\boldsymbol{m'},z)},$ let $\mathbb{F}_{Q}$ be an extension of $\mathbb{F}_{q},$ such that $Q > degree(g_n) + degree(g_d).$ As $\mathbb{F}_q \subset \mathbb{F}_Q,$ we can view the coefficients $\boldsymbol{m'}$ to be elements of $\mathbb{F}_Q,$ which we shall now refer to as $\boldsymbol{m'_Q}.$  We now choose some $z_{_Q}\in \mathbb{F}_Q$ such that $g_n(\boldsymbol{m'_Q},z)|_{z=z_{_Q}} \neq 0,$ and $g_d(\boldsymbol{m'_Q},z)|_{z=z_{_Q}} \neq 0.$ Such a choice is possible because the polynomial $g_n(\boldsymbol{m'_Q},z)g_d(\boldsymbol{m'_Q},z)$ can have at most $degree(g_n) + degree(g_d)$ zeros in $\mathbb{F}_Q.$ Therefore, with $z=z_{_Q},$ we have a well-defined network code for ${\cal G}_{inst}$ with $g \neq 0,$ satisfying the invertibility condition in ${\cal G}_{inst}$. Let the set of local encoding coefficients obtained for ${\cal G}_{inst}$ by assuming $z=z_{_Q}$ be $\boldsymbol{m''_Q}.$

As for the zero-interference conditions, $f_1(\boldsymbol{m'_Q},z)=f_2(\boldsymbol{m'_Q},z)=...=f_K(\boldsymbol{m'_Q},z)$ all being zero polynomials implies that any choice of $z$ does not alter their value. Therefore the network code defined by $\boldsymbol{m''_Q}$ is a feasible network code for ${\cal G}_{inst}({\cal V},{\cal E},{\cal S},{\cal T},{\cal C}).$ This completes the proof.
\end{IEEEproof}
\section{Proof of Corollary \ref{corollaryconstruction}}
\label{corollaryconstructionproof}
\begin{IEEEproof}
Let $\boldsymbol{m'}$ be a set of local encoding coefficients taking values from some field $\mathbb{F}_q$ which result in a feasible network code for ${\cal G}_{ud}$, satisfying the invertibility and zero-interference conditions. Let the resulting network transfer matrix of any particular sink $t\in \cal T$ in ${\cal G}_{ud}$ be  $M_t(z)$ and let $M'_t(z)$ be the $h_t \times h_t$ submatrix of $M_t(z)$ which involves those rows of $M_t(z)$ that correspond to the information sequences to be obtained by sink $t.$ Although $M'_t(z)$ involves rational functions in $z,$ the denominators can be factored out to obtain a matrix of the form $M'_t(z)=(a(z))^{-1}M''_t(z)$ where $a(z)\in\mathbb{F}_q[z]$ and $M''_t(z)$ is a matrix consisting of bounded-degree polynomials in $z.$ It is known (see \cite{MuS}, for example) that the determinant of a polynomial matrix of size $k\times k$ and degree at $d$ can be calculated with complexity $O(k^3d^2).$ The determinant of $M'_t(z)$ can thus be calculated with polynomial complexity. Thus, following the notations from the proof of Proposition \ref{prop2}, it is therefore clear that the product $g(\boldsymbol{m'},z)= \frac{g_n(\boldsymbol{m'},z)}{g_d(\boldsymbol{m'},z)}$ of the determinants of all the sinks can also be calculated in polynomial-time.

The next step in the construction of a feasible network code for ${\cal G}_{inst}$ according to the proof of Proposition \ref{prop2} is to pick a value $z_{_Q}$ from a large enough field $\mathbb{F}_{Q}$ so that $g(\boldsymbol{m'_Q},z_{_Q})\neq 0.$ Finding such a $z_{_Q}$ involves at most $degree(g_n) + degree(g_d)$ evaluations of the polynomial $g_ng_d.$ Such polynomial evaluations can be performed with complexity linear in the degree of the polynomial concerned (see \cite{BoM}, for example). Therefore, identifying an appropriate $z_Q$ takes $O\left(\left(degree(g_n)+degree(g_d)\right)^2 \right)$ operations over the field concerned. 

Once such a $z_{_Q}$ has been identified, the local encoding coefficients $\boldsymbol{m''_Q}$ for ${\cal G}_{inst}$ can be obtained  can be obtained by evaluating the local encoding coefficients $\boldsymbol{m'_Q}$ of ${\cal G}_{ud}$ at $z=z_{_Q}.$ All the elements of $\boldsymbol{m'_Q}$ are rational functions and there are at most $|{\cal E}|^2$ of them, thus the total complexity involved in these evaluations is also polynomial. Once these evaluations have been obtained, we have a feasible network code for ${\cal G}_{inst}.$ The complexity of obtaining such a feasible network code for ${\cal G}_{inst}$ at each step has been polynomial including that of the obtaining a feasible network code for ${\cal G}_{ud},$ as assumed. This proves the corollary.
\end{IEEEproof}
\section{Proof of Proposition \ref{prop3}}
\label{prop3proof}
\begin{IEEEproof}
Consider a feasible network coding solution from $\cal U$ for ${\cal G}_{ud}.$ Because of the conditions on the topology of the network, the columns of the $h \times h_t$ (following the notations in Section \ref{sec2}) network transfer matrix $M_t(z)$ of a sink $t \in \cal T$ are of the form 
\begin{equation}
\label{eqn6}
\boldsymbol{M_{i,t}}(z)=z^{a_i}\boldsymbol{M_{i,t}}, 1 \leq i \leq h_t
\end{equation}
where $\boldsymbol{M_{i,t}}(z)$ is the $i^{th}$ column of $M_t(z),$ $\boldsymbol{M_{i,t}} \in \mathbb{F}_q^{h}$ and $a_i \in \mathbb{Z}^+.$ We also have $M_t=M_t(z)|_{z=1},$ the network transfer matrix of the sink $t$ in ${\cal G}_{inst},$ the $i^{th}$ column of which is $\boldsymbol{M_{i,t}}.$ Let $M'_t(z)$ be the $h_t\times h_t$ submatrix of $M_t(z),$ involving those rows (say, those indexed by $\left\{j_i \in \left\{1,2,3,...,h\right\}:1 \leq i \leq h_t \right\}$) of $M_t(z)$ which correspond to the information sequences that need to be inverted at $t,$ and let $M'_t$ be the corresponding matrix for ${\cal G}_{inst}.$ Because of (\ref{eqn6}), the determinant of $M'_t(z)$ is of the form
\begin{equation}
\label{eqn8}
det\left(M'_t(z)\right)=det\left(M'_t\right)\prod_{i=1}^{h_t}z^{a_{j_i}}.
\end{equation}
Thus, if $det\left(M'_t(z)\right) \neq 0,$ then $det\left(M'_t\right)\neq 0,$ which means $M'_t$ is invertible. Also, any zero element of $M_t(z)$ is also zero in $M_t.$ As the choice of the sink $t$ was arbitrary, both the invertibility and the zero-interference conditions are satisfied for all sinks in ${\cal G}_{inst},$ thus proving (A).

We now prove (B). Suppose there is a feasible solution in place for ${\cal G}_{inst}.$ Because of the condition on the network topology, the network transfer matrices in ${\cal G}_{ud}$ is of the form (\ref{eqn6}). Then, by (\ref{eqn8}) the corresponding invertible $h_t \times h_t$ submatrix $M'_t(z)$ of the network transfer matrix  $M_t(z)$ of sink $t$ in ${\cal G}_{ud}$ with the same local encoding coefficients as ${\cal G}_{inst}$ has a non-zero determinant and is thus full-rank. Thus the invertibility conditions for sink $t$ are carried over to ${\cal G}_{ud}.$ To prove the zero-interference conditions, suppose $M_{i,j,t}$ is the element $(i,j)$ of $M_t$ which is zero. Then the corresponding element of $M_t(z),$ $M_{i,j,t}(z),$ is such that $M_{i,j,t}(z)=0,$ or $M_{i,j,t}(z)\neq 0$ with $(z-1)|M_{i,j,t}(z)$ (as $M_t(z)|_{z=1}=M_t$). However, because of (\ref{eqn6}), $M_{i,j,t}(z)=z^{a_i}M_{i,j,t}$ which means that $(z-1)\nmid M_{i,j,t}(z).$ Thus $M_{i,j,t}(z)=0.$ The zero-interference conditions are also satisfied for $t$ in ${\cal G}_{ud}.$ Again, as the choice of sink $t$ was arbitrary, any solution for ${\cal G}_{inst}$ is also a solution for ${\cal G}_{ud}.$

To prove (C), we first note that, by (B), the minimum field size requirement for a feasible solution for ${\cal G}_{ud}$ is not larger than that of ${\cal G}_{inst}.$ Also, by (A), any solution for ${\cal G}_{ud}$ from any non-empty ${\cal U}_q$ is feasible for ${\cal G}_{inst},$ which holds for $q=q_{min}$ too. This fact along with (B) proves (C).
\end{IEEEproof}
\section{Proof of Proposition \ref{prop4}}
\label{prop4proof}
\begin{IEEEproof}
Throughout this proof, we assume that the network we are working with is ${\cal G}^*({\cal V}^*,{\cal E}^*),$ the subnetwork of ${\cal G}$ consisting only of the nodes and edges on the $h_s$ edge-disjoint paths from the source $s$ to each sink $t\in \cal T$. Just as Lemma \ref{lemma2} together with Lemma \ref{lemma1} justifies the maintenance of the rank of the matrices $B_t: t\in {\cal T}$ in every step of the LIF algorithm, we prove this proposition by showing a variant of Lemma \ref{lemma2} which will maintain the rank of the matrices $B_t: t\in {\cal T}$ according to the delay-and-code schemes. 

Let $e$ be the edge whose global encoding vector is to be decided in the current step of the non-uniform delay-and-code LIF algorithm, and let $v=tail(e)$. We have sets of ordered pairs (elements from $\mathbb{F}_q(z)^{h_s}\times \mathbb{F}_q(z)^{h_s}$) as in Lemma \ref{lemma2}, 
\begin{align}
\nonumber
S(e):&=\left\{(\boldsymbol{x_i},\boldsymbol{y_i}):i \leq i \leq n\right\}\\
\label{eqn14}
&=\left\{(\boldsymbol{b}(f^t_{\leftarrow}(e)),\boldsymbol{a_t}(f^t_{\leftarrow}(e))):t\in T(e) \right\},
\end{align}
such that $\boldsymbol{x_i}.\boldsymbol{y_i} \neq 0, 1\leq i \leq n,$ $n$ being the cardinality of the set in the RHS.   We seek to iteratively construct the vectors $\boldsymbol{u_1},\boldsymbol{u_2},...,\boldsymbol{u_n}$ (each $\boldsymbol{u_i} \in \mathbb{F}_q(z)^{h_s}$) such that the following conditions hold for each $i, 1\leq i \leq n$.
\begin{enumerate}
\item $\boldsymbol{u_i}$ is some delay-and-code based linear combination of the vectors $\left\{\boldsymbol{x_k}: 1\leq k \leq i\right\}.$ 
\item For any $j$ such that $1\leq j \leq i, \boldsymbol{u_i}.\boldsymbol{y_j} \neq 0.$
\end{enumerate}
If such vectors can be found, then we fix $\boldsymbol{b}(e)=\boldsymbol{u_n}$ as the global encoding vector of $e$ as it can be seen using Lemma \ref{lemma1} that such a choice preserves the necessary requirements for the current step of the non-uniform delay-and-code LIF algorithm. The vectors $\boldsymbol{u_1},\boldsymbol{u_2},...,\boldsymbol{u_n}$ are constructed as follows. 

Let $\boldsymbol{u_1} = \alpha_1\boldsymbol{x_1},$ where $\alpha_1 \in \mathbb{F}_2$ and $\alpha_1 \neq 0$ as we need $\boldsymbol{u_1}.\boldsymbol{y_1} \neq 0.$ Now suppose for some $i, 1\leq i \leq n-1,$ we have $\boldsymbol{u_i}$ such that $\boldsymbol{u_i}.\boldsymbol{y_j} \neq 0 ~\forall 1 \leq j \leq i.$  Then we will show that we can get $\boldsymbol{u_{i+1}}$ such that $\boldsymbol{u_{i+1}}.\boldsymbol{y_j} \neq 0 ~\forall 1 \leq j \leq i+1,$ as long as the total number of memory elements present at each node for each incoming edge is at least $\left(|{\cal T}|-1\right).$ 

Suppose $\boldsymbol{u_i}.\boldsymbol{y_{i+1}}\neq 0.$ Then we choose $\boldsymbol{u_{i+1}}=\boldsymbol{u_i},$ which then satisfies our requirements. Else, we choose 
\begin{equation}
\label{eqn12}
\boldsymbol{u_{i+1}}=\boldsymbol{u_i}+\alpha_{i+1}z^{\beta_{i+1}}\boldsymbol{x_{i+1}}.
\end{equation}
Again, we have $\alpha_{i+1} \in \mathbb{F}_2$ and $\alpha_{i+1} \neq 0$ as we want $\boldsymbol{u_{i+1}}.\boldsymbol{y_{i+1}} \neq 0$. $\beta_{i+1}\in\mathbb{Z}^{\geq 0}$ is the number of memory elements used to delay the symbols on that particular incoming edge. Thus $\boldsymbol{u_{i+1}}=\boldsymbol{u_i} + z^{\beta_{i+1}}\boldsymbol{x_{i+1}}.$ 

Now suppose for some choice of $\beta_{i+1}=\beta$ and for some $j, 1 \leq j \leq i,$ we have  $\boldsymbol{u_{i+1}}.\boldsymbol{y_j}=0,$ i.e., 
\begin{equation}
\label{eqn13}
\boldsymbol{u_{i}}.\boldsymbol{y_j} = - z^{\beta}\left(\boldsymbol{x_{i+1}}.\boldsymbol{y_j}\right),
\end{equation}
then $\beta$ is not a valid choice for $\beta_{i+1},$ as (\ref{eqn13}) should not hold for any $j, 1\leq j\leq i.$ Note that there are at most $i$ choices for $j$ at which (\ref{eqn13}) will hold. There are therefore at most $i$ choices for $\beta_{i+1}$ that cannot be used. As $1 \leq i \leq n-1,$ if we have at least $n$ choices for $\beta_{i+1},$ then we can always choose one value such that (\ref{eqn13}) does not hold for any $1\leq j \leq i$ for any given $i, 1\leq i\leq n-1.$ With $|{\cal T}|-1$ memory elements for each incoming edge at node $v,$ we have $|{\cal T}|$ choices for any particular $\beta_{i+1}.$ This, coupled with Lemma \ref{lemma1} and the fact that $n \leq |{\cal T}|$ ensures that a non-uniform delay-and-code scheme can be constructed for the given multicast problem.
\end{IEEEproof}

\section{Proof of Proposition \ref{prop5}}
\label{prop5proof}
\begin{IEEEproof}
As in Proposition \ref{prop4}, we prove the proposition using a variant of the proof for Lemma \ref{lemma2}. We again assume that the network we are working with is ${\cal G}^*({\cal V}^*,{\cal E}^*),$ the subnetwork of ${\cal G}$ consisting only of the nodes and edges on the $h_s$ edge-disjoint paths from the source $s$ to each sink $t\in \cal T$. We follow an ancestral ordering which enables us to process all outgoing edges of a particular intermediate node before moving to the next. As the uniform delay-and-code technique is defined only for the intermediate nodes, we use the non-uniform delay-and-code technique at the source node to preserve the ranks of the matrices $B_t$ for each sink $t,$ until all the edges in $\Gamma_O(s)$ have been processed.

Let $v$ be the intermediate node whose outgoing edges are to be processed together. As in the proof of Proposition \ref{prop4}, for each $e \in \Gamma_O(v),$ we have the set $S(e)$ as defined in (\ref{eqn14}), and we seek to iteratively construct $\boldsymbol{u_1},\boldsymbol{u_2},...,\boldsymbol{u_n},$ such that conditions $1)$ and $2)$ (as in the proof of Proposition \ref{prop4}) are satisfied. Also, (\ref{eqn11}) needs to be satisfied because we seek to design a uniform delay-and-code.

In the process of choosing each $\boldsymbol{u_i}$ for any particular $e\in\Gamma_O(v),$ the local encoding coefficient $m_{e,p}$ and the delay $a_{e,p}$ ($m_{e,p}$ and $a_{e,p}$ are as in (\ref{eqn10})) corresponding to $e$ and some $p\in\Gamma_I(v)$ has to be chosen. Based on the arguments developed in the proof of Proposition \ref{prop4}, the choices for $m_{e,p}$ and $a_{e,p}$ are restricted.  For any particular $p\in \Gamma_I(v)$ and for any edge $e\in \Gamma_O(v)$ for which $m_{e,p}\neq 0,$  it was shown that there are at most $|{\cal T}|-1$ choices that are not allowed for $a_{e,p}$ in the non-uniform delay-and-code case. 

For the uniform delay-and-code case, we need $a_{e,p}=a_p, \forall e\in\Gamma_O(v)$ such that $m_{e,p}\neq 0.$ Thus the number of choices of $a_p$ that cannot be allowed for any edge $p\in \Gamma_I(v)$ is at most $\delta\left(|{\cal T}|-1\right),$ in the case of $m_{e,p}\neq 0~ \forall e\in\Gamma_O(v).$ If $m_{e,p}=0$ for any $e\in\Gamma_O(v),$ this number of disallowed choices for $a_{e,p}$ can only reduce. Therefore if every node in the given network has $\delta\left(|{\cal T}|-1\right)$ memory elements for each incoming edge, then there exists at least one choice for $a_p$ such that the conditions on the invertibility of the $B_t$ matrices (again by invoking Lemma \ref{lemma1}) and the uniformity of the delay-and-code scheme given by (\ref{eqn11}) are satisfied. 

Note that the proof hinges on the fact that the network has node-disjoint paths, as we have seen in Subsection \ref{delayandcode}, that a feasible uniform delay-and-code scheme might not be possible to design in a general network with only edge-disjoint paths. This concludes the proof.
\end{IEEEproof} 
\section{Proof of Corollary \ref{cor4}}
\label{cor4proof}
\begin{IEEEproof}
Proposition \ref{prop5} shows that a feasible network code based on the uniform delay-and-code scheme can be constructed for a multicast situation on networks with node-disjoint paths. Note that given such a network, all that the uniform delay-and-code scheme effectively does is to introduce different delays on the incoming edges. Because of the uniformity of the delay-and-code scheme, i.e., the formulation given by (\ref{eqn11}), the $a_{p}$ memory elements used for each edge $p\in\Gamma_I(v)$ at some intermediate node $v$ can be viewed as $a_{p}$ additional delays on edge $p,$ or equivalently as additional forwarding nodes with $a_p$ forwarding edges. In other words, given an acyclic network ${\cal G}$ with multicast demands and node-disjoint paths, a feasible uniform delay-and-code scheme was obtained for the unit delay network ${\cal G}_{ud}$. Then the unit-delay network ${\cal G}_{ud}$ along with the uniform delay-and-code network code naturally invokes the unit-delay network $\tilde{{\cal G}}_{ud}$ on which there exists a feasible network code over $\mathbb{F}_2,$ by using the equivalence between the memory elements and delays. 
\end{IEEEproof}
\end{document}